\documentclass{amsart}
\usepackage{amsfonts}
\usepackage{amssymb}
\usepackage{hyperref}

\newtheorem{theorem}{Theorem}[section]
\newtheorem{lemma}[theorem]{Lemma}
\theoremstyle{definition}

\newtheorem{corollary}[theorem]{Corollary}

\theoremstyle{remark}
\newtheorem{remark}[theorem]{Remark}
\numberwithin{equation}{section}

\newcommand{\BR}{{\mathbb R}}
\newcommand{\BC}{{\mathbb C}}
\newcommand{\BN}{{\mathbb N}}
\newcommand{\BZ}{{\mathbb Z}}
\newcommand{\Qh}{\left(-\frac{\pi}{h},\frac{\pi}{h}\right]^n}

\renewcommand{\a}{\textbf{a}}
\renewcommand{\b}{\textbf{b}}

\newcommand{\f}{\textbf{f}}

\newcommand{\g}{\textbf{g}}

\newcommand{\e}{{\bf e}}

\newcommand{\KH}{{\bf K}_H}

\newcommand{\FH}{{\bf F}_H}

\newcommand{\cl}{C \kern -0.1em \ell}

\begin{document}
\title{Time-changed Dirac-Fokker-Planck equations on the lattice}
\author{N.~Faustino}
\address{CMCC, Universidade Federal do ABC, 09210--580, Santo Andr\'e, SP,
	Brazil}
\email{\href{mailto:nelson.faustino@ufabc.edu.br}{nelson.faustino@ufabc.edu.br}~|~\href{mailto:nelson.faustino@ymail.com}{nelson.faustino@ymail.com}}
\date{\today}

\begin{abstract}
 A time-changed discretization for the Dirac equation is proposed. More precisely, we consider a Dirac equation with discrete space and continuous time perturbed by a time-dependent diffusion term $\sigma^2Ht^{2H-1}$ that seamlessly describes a latticizing version of the time-changed Fokker-Planck equation carrying the Hurst parameter $0<H<1$. Our model problem formulated on the space-time lattice $\BR_{h,\alpha}^n\times [0,\infty)$ ($h>0$ and $0<\alpha<\frac{1}{2}$) preserves the main features of the Dirac-K\"ahler type discretization over the space-time lattice $h\BZ^n\times [0,\infty)$ in case of $\alpha,H \rightarrow 0$, and encompasses a regularization of Wilson's approach [Physical review D, 10(8), 2445, 1974] for values of $H$ in the range $0<H\leq \frac{1}{2}$ (limit condition $\alpha \rightarrow \frac{1}{2}$).
 
The main focus here is the representation of the solutions by means of discrete convolution formulae involving a kernel function encoded by (unnormalized) Hartman-Watson distributions -- ubiquitous on stochastic processes of Bessel type -- and the solutions of a semi-discrete equation of Klein-Gordon type. Namely, on our main construction the ansatz function $\widehat{\varPsi}_H(y)$ appearing on the discrete convolution representation may be rewritten as a Mellin convolution type integral involving the solutions $\varPsi(x,t|p)$ of a semi-discrete equation of Klein-Gordon type and a L\'evy one-sided distribution $L_H(u)$ in disguise. Interesting enough, by employing Mellin-Barnes integral representations it turns out that the underlying solutions of Klein-Gordon type may be represented through generalized Wright functions of type ${~}_1\Psi_1$, that converge uniformly in case that the quantity $\alpha+\frac{1}{2}$ may be regarded as an lower estimate for the Hurst parameter in the superdiffusive case (that is, if $\alpha+\frac{1}{2}\leq H<1$).
\end{abstract}

\maketitle

\section{Introduction}\label{IntroductionSection}

\subsection{General overview}
Apart the discretization of the Klein-Gordon equation, the discretization of Dirac-type equations is one of the most deeply study equations in lattice gauge theories due to its implications on the formulation of \textit{Quantum Electrodynamics} (QED) and \textit{Quantum Chromodynamics} (QCD) on space-time lattices (see, for instance, \cite[Chapters 4 \& 5]{MontvayMunster97}). And as it well-known from Wilson, Kogut-Susskind and Rabin fundamental papers (cf.~\cite{Wilson74,KS75,Rabin82}),
the latticizing versions of such equations does not lead, in general, to its continuous counterparts in the continuum theory, due to the presence of doubler fermions-- the so-called \textit{lattice fermion doubling phenomena}, characterized in detail in Nielsen-Ninomiya's paper \cite{NN81} (see also \cite[subsection 4.4]{MontvayMunster97}). 

The \textit{lattice fermion doubling} holds because the momentum space carrying a lattice with meshwidth proportional to $h>0$ -- the so-called \textit{Brillouin zone} -- has the topology of the $n-$torus $\BR^n/\frac{2\pi}{h}\BZ^n$. While in Wilson's approach \cite{Wilson74} the doubler fermions were removed by adding a cut-off term depending upon the discretization of the Laplace operator $\Delta$ on the lattice $h\BZ^n$ in \cite{Rabin82} it was shown that a staggered fermionic version of Nielsen-Ninomiya's approach, initially proposed on the seminal paper \cite{KS75} of Kogut and Susskind, turn out to be linked with the discretization of the Dirac-K\"ahler operator $d-\delta$ (cf.~\cite{Rabin82}). We refer to \cite[subsection 4.3]{MontvayMunster97} for an overview of Kogut-Susskind's approach \cite{KS75}, to \cite{Sushch14} for a detailed application of Dirac-K\"ahler formalism toward homology theory and \cite{Faustino16,FaustinoMMAS17} for a multivector calculus perspective on the lattice $h\BZ^n$.

Mainly influenced by the approaches considered in \cite{FaustinoKS07,RSKS10}, the discretization of Dirac equations \textit{\`a la Dirac-K\"ahler} (cf.~\cite{Rabin82}) has been widely used on the last decade to develop further perspectives on the field of discrete harmonic analysis. We refer to \cite{BRS12} for a multivector formulation of discrete Fourier analysis, to \cite{BaaskeBRS14} for applications on the theory of discrete heat semigroups and to \cite{CKKS14} for a higher-dimensional extension of the theory of discrete Hardy spaces.

In author's recent paper \cite{FaustinoRelativistic18} the construction of discretizations for the Dirac operator was reformulated from a pseudo-differential calculus perspective. The idea was to relate directly the construction of discrete Dirac operators to the theory of discrete distributions (see \cite[Section 2.]{FaustinoRelativistic18}) with the aid of representation of its Fourier multipliers on the momentum space $\Qh$. As a result it was possible to obtain not only a discrete Dirac operator, as in \cite{FaustinoKS07,RSKS10,BRS12,Faustino16,FaustinoMMAS17}, but a family of discrete Dirac operators $D_{h,\alpha}$ carrying the fractional parameter $0<\alpha<\frac{1}{2}$. Each one turn out to be connected to the 'fractional' lattice $\BR_{h,\alpha}^n:=(1-\alpha)h\BZ^n\oplus \alpha h\BZ^n$.

Summing up, this approach combines the discrete Fourier analysis framework proposed by G\"urlebeck and Spr\"o\ss ig in \cite[Chapter 5]{GuerlebeckSproessig97}  with some abstract results on discrete distributions studied in depth by Ruzhansky and Turunen in \cite[Part II]{RuzhanskyT10}. And in contrast with \cite{BRS12,BaaskeBRS14}, the underlying spaces of discrete distributions turn out to be linked with the topology of the $n-$torus, through the canonical isomorphism $\mathbb{R}^n/\frac{2\pi}{h}\mathbb{Z}^n\cong \Qh$ (see, for instance, \cite[Section 5]{BRS12} and \cite[Section 5]{BaaskeBRS14} for further comparisons) in a way that it is possible to have a physical interpretation for the \textit{lattice fermion doubling phenomena} which does not conflict with the ladder structure of the Clifford algebra (cf.~\cite[Section 2]{Faustino16}), nor in case where additional symmetries are involved.

Unfortunately, one cannot say the same for the Wilson's approach \cite{Wilson74} in the \textit{continuum limit}, due to the following facts: the chiral symmetries are not recovered and the additional fermion doublers do not remain on the spectrum (we refer to \cite[Section 4.2]{MontvayMunster97} for more details). From a mathematical perspective, that roughly means that Nielsen-Nimomiya's no-go result \cite{NN81} was evaded since, contrary to \cite{KS75,Rabin82}, the additional lattice fermion doublers appearing on Wilson's formulation do not depend on the signature of the Clifford algebra, although the Green's function carrying Wilson propagator behaves as the Green's function carrying a free fermion field (that is, the fundamental solution of the continuous Dirac operator). 

Following Mandelbrot and Van Ness \cite{MandelNess68} seminal paper on \textit{fractional Brownian motion} (fBM)  (see also \cite[subsection 7.6]{MeerSik11}) which has been
gained widespread attention on the stochastic analysis community, mainly due to Hairer's fundamental paper \cite{Hairer05},
it becomes natural to inquire if the discrete Dirac equation \textit{a la Wilson} proposed in \cite{Wilson74} can be interpreted as a time-changed stochastic process. 

Of course, the idea of describing physical models depending on phase transitions through fBM is indeed very old, as one may notice e.g. on Wilson's quotation during his Nobel Prize lecture \cite{Wilson82} (that may be found in \cite[p.~124]{Wilson82}):
\begin{quote}
	{\it There is a murky connection between scaling ideas in critical phenomena and
		Mandelbrot’s “fractals” theory - a theory of scaling of irregular geometrical
		structures (such as coastlines).}
\end{quote}

Nevertheless, due to the ongoing research interest on discrete multivector structures one may view this work as a first step to identify further directions towards a stochastic perspective, with the aim of enrich the framework that has been developed in the series of papers \cite{FaustinoKS07,RSKS10,BRS12,BaaskeBRS14,CKKS14,Sushch14,Faustino16,FaustinoMMAS17,FaustinoRelativistic18}.

\subsection{Statement of the model problem}

The model problem under consideration -- that it will be coined here and elsewhere as \textit{time-changed Dirac-Fokker-Planck} (DFP) equation on the lattice -- is strongly motivated by the recent surge of interest on the theory of time-changed Fokker-Planck equations in \textit{continuum} (cf. \cite{HKU11,HRKU11}) and by the ongoing promissing applications of such laticizing models toward stochastic discretization (cf.~\cite{Tarasov14}). Other additional motivations may be found on Hairer's paper \cite{Hairer05} -- devoted to an exploitation of Mandelbrot-Van Ness's approach \cite{MandelNess68} to stochastic PDEs -- and on Hairer et al. papers \cite{HairerMassWeber14,HairerMat18} -- centered on rigorous goal-oriented formulations for discrete counterparts for (non-linear) stochastic PDEs. Its structure is organized as follows: 
\begin{itemize}
	\item In Section \ref{Preliminaries} we introduce the framework we will work with, namely the main ingredients and features of Clifford algebras, discrete Fourier analysis and of the representations of discrete Dirac and discrete Laplacians on the 'fractional' lattice $\BR_{h,\alpha}^n:=(1-\alpha)h\BZ^n\oplus \alpha h \BZ^n$ ($h>0$ and $0<\alpha<\frac{1}{2}$).
	\item In Section \ref{TimeChangedFPDSection} we propose a possible a time-changed version for the Dirac equation on the space-time lattice -- time-changed DFP for short -- which is akin to a regularization of Wilson-Dirac equation in case where $0<H\leq \frac{1}{2}$. It will be depicted, in particular, some connections between the fundamental solution of the semi-discrete heat operator $\partial_t-\Delta_h$ and stochastic processes of Bessel type.
	\item Section \ref{MainSection} will be devoted to the main results of the paper, mainly to the representation of time-changed DFP equation as a discrete convolution between the fundamental solution of the semi-discrete heat equation and the solution of a semi-discrete equation of Klein-Gordon type. Moreover, the Mellin-Barnes framework will be considered to show that the underlying solution of the Klein-Gordon equation admits analytic representations, involving \textit{generalized Wright functions} of type ${~}_1\Psi_1$ in the superdiffusive case ($\frac{1}{2}+\alpha\leq H<1$, with $0<\alpha<\frac{1}{2}$).
\end{itemize}

Throughout this paper, the time-changed DFP equation on the space-time lattice $\BR^n_{h,\alpha}\times [0,\infty)$ introduced in subsection \ref{ModelProblemSub} is far from being a simple second-order perturbation of the discrete Dirac equation studied in author's recent paper \cite{FaustinoRelativistic18}. In the limit $h\rightarrow 0$, it corresponds to a finite difference discretization of a Gaussian process $\displaystyle X=\sum_{j=1}^n \e_j X_j$ carrying a set of \textit{independent and identically distributed} (i.i.d) random variables $X_1,X_2,\ldots,X_n$ with variance $\sigma^2_H(t)=\sigma^2t^{2H}$ (see e.g. \cite[Proposition 1. \& Remark 2.]{HRKU11}). Herein, the geometric calculus nature of Clifford algebras highlighted in the series of books (see e.g. \cite{GuerlebeckSproessig97,VazRoldao16}) allows us to use the discretizations for the Dirac operator considered previously in  \cite{FaustinoKS07,RSKS10,Sushch14,Faustino16,FaustinoMMAS17} to develop more robust algebraic tools to go from one-dimensional fractional diffusion models to higher dimensional ones (cf.~\cite[Chapter 6]{MeerSik11}) such as the time-changed regularization of the Wilson-Dirac equation \cite{Wilson74} highlighted on \textbf{Remark \ref{WilsonRemark}} and \textbf{Remark \ref{fBMRemark}}.

 The main results treated in section \ref{MainSection} are essentially \textbf{Theorem \ref{mainResultDFP}} and \textbf{Corollary \ref{LevyRepCorollary}}: In the proof of \textbf{Theorem \ref{mainResultDFP}} we tackle the problem of representing the solutions within the framework introduced in subsections \ref{DiscreteFourierAnalysis}, \ref{discreteOpSub} and \ref{StochasticSubsection} whereas in the proof of \textbf{Corollary \ref{LevyRepCorollary}} we show that if we known {\it a-priori} the solution of the semi-discrete Klein-Gordon equation (\ref{KleinGordonParametrix}) (see \textbf{Theorem \ref{KleinGordonAnsatz}}), then the solution $\varPsi(x,t|p)$ of the time-changed DFP equation (\ref{DiracFokkerPlanck}) can be neatly represented as a Laplace type integral that encompasses $\varPsi(x,t|p)$ and the stable one-side L\'evy distributions $L_H(u)$ depending on the Hurst parameter $0<H<1$.
 
 For our purposes (time-changed equation depending on the Hurst parameter $0<H<1$) it becomes relevant to consider, as in author's recent paper \cite{FaustinoBayesian17}, the generalized Wright functions ${~}_p\Psi_q$
 to encompass the stable one-side L\'evy distributions $L_H(u)$ of order $0<H<1$ appearing on subsection \ref{DFPKleinGordonsub} and the Fourier multipliers 
 \begin{eqnarray*}
 \cos\left(\mu t\sqrt{d_h(\xi)^2}\right)&\mbox{and}&\dfrac{\sin(\mu t\sqrt{d_h(\xi)^2})}{\sqrt{d_h(\xi)^2}}
 \end{eqnarray*}
appearing on the proof of {\bf Lemma \ref{expitzhLemma}}.
 For the sake of the reader's convenience we will outline some definitions required in Appendix section \ref{FractionalCalcAppendix} required for the proof of \textbf{Corollary \ref{LevyRepCorollary}}, and later, for the proof of \textbf{Theorem \ref{MellinBarnesTheorem}} in subsection \ref{GeneralizedWrightSubsection}.

Similarly to \cite{BaaskeBRS14,CiaurriGRTV17} our approach relies heavily on the representation of the fundamental solution of semi-discrete heat operator $\partial_t-\Delta_h$ in terms of modified Bessel functions of the first kind (see also \cite[subsection 4.2]{FaustinoRelativistic18}), in view of its major utility in treating initial value problems of Cauchy type through semigroup analysis. Loosely speaking, we have shown that the same scheme also works to represent the fundamental solution of its time-changed counterpart $\partial_t-H\sigma^2t^{2H-1}\Delta_h$ (zero-drift case of equation (\ref{DiracFokkerPlanck})).

The role of the modified Bessel functions and alike are indeed compelling and undisputed in the series of papers on the literature (see e.g. \cite{CiaurriGRTV17} and the references therein). In particular, its interplay with the discrete heat semigroup in the setting of discrete Clifford analysis has already been fully answered by Baaske et al. \cite{BaaskeBRS14} (see also \cite[subsection 4.2]{FaustinoRelativistic18}). 
With the incorporation of a less known interpretation for the fundamental solution of $\partial_t-\Delta_h$ in the modeling of Brownian motion through a stochastic process of Bessel type, as formerly outlined by Yor \cite{Yor80} (see also \cite{BorodinSalm12} and the references given there), we are able to produce a stochastic interpretation that do precisely fit our needs, 
as briefly depicted throughout subsections \ref{StochasticSubsection} and \ref{DFPKleinGordonsub}.

\section{Preliminaries}\label{Preliminaries}

\subsection{Clifford algebra setup}

Following the standard definitions considered in the book \cite{VazRoldao16} and the standard notations already considered in \cite{Faustino16,FaustinoMMAS17,FaustinoRelativistic18}, we will introduce the Clifford algebra of signature $(n,n)$ in the following way:

We will denote by $\cl_{n,n}$ the Clifford endowed by the Minkowski space-time $\BR^{n,n}$, and by $\e_1,\e_2,\ldots,\e_n$ , 
$\e_{n+1},\e_{n+2}\,\ldots,\e_{2n}$ the underlying basis of $\BR^{n,n}$. Herein we assume that $\cl_{n,n}$ is generated by the set of graded anti-commuting relations
\begin{eqnarray}
\label{CliffordBasis}
\begin{array}{lll}
\e_j \e_k+ \e_k \e_j=-2\delta_{jk}, & 1\leq j,k\leq n \\
\e_{j} \e_{n+k}+ \e_{n+k} \e_{j}=0, & 1\leq j,k\leq n\\
\e_{n+j} \e_{n+k}+ \e_{n+k} \e_{n+j}=2\delta_{jk}, & 1\leq j,k\leq
n.
\end{array}
\end{eqnarray}

Through the linear space isomorphism between $\cl_{n,n}$ and the exterior algebra $\bigwedge (\BR^{n,n})$ provided by the linear extension of the mapping
$\e_{j_1}\e_{j_2}\ldots \e_{j_r} \mapsto dx_{j_1}dx_{j_2}\ldots
dx_{j_r}$ ($1\leq j_1<j_2<\ldots<j_r\leq 2n$), it readily follows that the basis elements of $\cl_{n,n}$ consists on $r-$multivectors of the form $\e_{J}=\e_{j_1}\e_{j_2}\ldots
\e_{j_r}$ (cf.~\cite[Chapter 4]{VazRoldao16}). In case where $J=\varnothing$ (empty set)
we have $\e_\varnothing=1$ (basis for scalars).

For our main purposes we will make use of the embedding $\BR^{n,n}\subseteq \cl_{n,n}$ to represent, in particular, any $n-$tuple $(x_1,x_2\ldots,x_n)$ of $\BR^n$ 
by means of the linear combination $\displaystyle x=\sum_{j=1}^n x_j \e_j$ carrying the basis elements $\e_1,\e_2,\ldots,\e_n$ with signature $(0,n)$. In the same order of ideas, we will represent the associated translations $(x_1,x_2,\ldots, x_j\pm \varepsilon,\ldots,x_n)$
on lattices of the form $\varepsilon\BZ^n \subseteq \BR^n$
($\varepsilon>0$) as $x\pm \varepsilon\e_j$.

Now let
\begin{eqnarray*}
	\BR^n_{h,\alpha}:=(1-\alpha) h\BZ^n \oplus \alpha h\BZ^n,&\mbox{}~~~~h>0&\mbox{\&}~~~~0< \alpha< \frac{1}{2} 
\end{eqnarray*}
be a lattice of $\BR^n$ that contains $h\BZ^n$.

To properly introduce in \textbf{Section \ref{DiscreteFourierAnalysis}} a discrete Fourier transform carrying discrete multivector functions $\g:\BR^n_{\alpha,h} \rightarrow \BC\otimes\cl_{n,n}$ and $\f:\BR^n_{\alpha,h} \times [0,\infty) \rightarrow \BC\otimes\cl_{n,n}$ represented through one of the following ansatz ($\e_{J}=\e_{j_1}\e_{j_2}\ldots \e_{j_r}$)
\begin{eqnarray*}
	\g(x)=\displaystyle \sum_{r=0}^n\sum_{|J|=r} g_J(x) \e_J, & \mbox{with}& g_J:\BR^n_{\alpha,h} \rightarrow \BC \\
	\f(x,t)=\displaystyle \sum_{r=0}^n\sum_{|J|=r} f_J(x,t) \e_J, &\mbox{with}& f_J:\BR^n_{\alpha,h} \times [0,\infty) \rightarrow \BC
\end{eqnarray*}
we need to consider the $\dag-${\it
	conjugation} operation $\a \mapsto\a^\dag$ on the \textit{complexified Clifford algebra} $\BC\otimes\cl_{n,n}$ defined as 
\begin{eqnarray}
\label{dagconjugation}
\begin{array}{lll}
(\a \b)^\dag=\b^\dag\a^\dag \\ (a \e_J)^\dag =\overline{a_J}~\e_{j_r}^\dag
\ldots \e_{j_2}^\dag\e_{j_1}^\dag~~~(1\leq j_1<j_2<\ldots<j_r\leq 2n) \\
\e_j^\dag=-\e_j~~~\mbox{and}~~~\e_{n+j}^\dag=\e_{n+j}~~~(1\leq j\leq
n)
\end{array}.
\end{eqnarray}

It is straightforward to infer from (\ref{dagconjugation}) that $\a^\dag\a=\a\a^\dag$ is non-negative so that the $\|\cdot \|-$norm endowed by \textit{complexified Clifford algebra} structure $\BC\otimes \cl_{n,n}$ is defined as $\| \a\|=\sqrt{\a^\dagger \a}$. In case where $\a$ belongs to $\BC\otimes \BR^{n,n}$, it then follows that the quantity $\|\a\|$ coincides with the standard norm of $\a$ on $\BC^{2n}$.

To avoid ambiguities throughout the manuscript we will use the bold notations  $\a,\b,\ldots,\f,\g,\ldots$ when we refer to Clifford numbers and/or multivector functions with membership in the \textit{complexified Clifford algebra} $\BC\otimes\cl_{n,n}$.

\subsection{Discrete Fourier Analysis}\label{DiscreteFourierAnalysis}

Let $\ell_2(\BR^n_{h,\alpha};\BC\otimes \cl_{n,n}):=\ell_2(\BR^n_{h,\alpha})\otimes \left(\BC\otimes \cl_{n,n}\right)$ denotes the \textit{right Hilbert module} endowed by the Clifford-valued sesquilinear form (cf.~\cite[p.~533]{FaustinoBayesian17})
\begin{eqnarray}
\label{lpInner} \langle \f(\cdot,t),\g(\cdot,t) \rangle_{h,\alpha}=\sum_{x\in \BR^n_{h,\alpha}}h^n~
\f(x,t)^\dag\g(x,t),
\end{eqnarray}
and let $\mathcal{S}(\BR^n_{h,\alpha};\cl_{n,n}):=\mathcal{S}(\BR^n_{h,\alpha})\otimes \left(\BC\otimes \cl_{n,n}\right)$ denote the space of \textit{rapidly decaying functions} $\f$ with values on $\BC\otimes\cl_{n,n}$, defined for any $\BR-$valued constant $M<\infty$ by the semi-norm condition
$$ \displaystyle \sup_{x \in \BR^n_{h,\alpha}} (1+\| x\|^2)^M~\| \f(x,t)\|<\infty.$$

Following {\it mutatis mutandis} \cite[Exercise 3.1.7]{RuzhanskyT10}, it is straighforward to see that the seminorm condition
$$ \displaystyle \sup_{x \in \BR^n_{h,\alpha}} (1+\| x\|^2)^{-M}~\| \g(x,t)\|<\infty$$
allows us to properly define the set of all \textit{continuous linear functionals} with membership in $\mathcal{S}(\BR^n_{h,\alpha};\BC\otimes \cl_{n,n})$ through the mapping $\f(\cdot,t) \mapsto \langle \f(\cdot,t),\g(\cdot,t)\rangle_{h,\alpha}$, whereby the family of distributions $\g(\cdot,t):\BR^n_{h,\alpha} \rightarrow \BC\otimes \cl_{n,n}$ (for every $t\in [0,\infty)$) belong to the multivector counterpart of the \textit{space of tempered distributions} on the lattice $\BR^n_{h,\alpha}$. This function space will be denoted here and elsewhere by $$\mathcal{S}'(\BR^n_{h,\alpha};\BC\otimes \cl_{n,n}):=\mathcal{S}'(\BR^n_{h,\alpha})\otimes\left(\BC\otimes \cl_{n,n}\right).$$  

Next, let us denote by $\Qh$ an 
$n-$dimensional representation of the $n-$torus $\BR^n/\frac{2\pi}{h}\BZ^n$ and by $$L_2\left(\Qh;\BC\otimes\cl_{n,n}\right):=L_2\left(\Qh\right)\otimes\left(\BC\otimes \cl_{n,n}\right)$$ the $\BC \otimes \cl_{n,n}-$\textit{Hilbert module} endowed by the sesquilinear form
\begin{eqnarray}
\label{BilinearFormQh}\langle \f(\cdot,t),\g(\cdot,t
)\rangle_{\Qh}= \int_{\Qh} \f(\xi,t)^\dag \g(\xi,t) d\xi.
\end{eqnarray}

 The \textit{discrete Fourier transform} of a function $\g(\cdot,t)$ with membership in $\ell_2(\BR^n_{h,\alpha};\BC\otimes \cl_{n,n})$ is defined as
\begin{eqnarray}
\label{discreteFh}
(\mathcal{F}_{h,\alpha} \g)(\xi,t)=\left\{\begin{array}{lll}
\displaystyle \frac{h^n}{\left(2\pi\right)^{\frac{n}{2}}}\displaystyle 
\sum_{x\in \BR^n_{h,\alpha}}\g(x,t)e^{i x \cdot \xi}, & \xi\in \Qh
\\ \ \\
0 , & \xi\in \BR^n \setminus \Qh
\end{array}\right..
\end{eqnarray}

As in \cite[subsection 5.2.1]{GuerlebeckSproessig97}, the \textit{discrete Fourier transform} yields the isometric isomorphism $$\mathcal{F}_{\alpha,h}:\ell_2(\BR^n_{h,\alpha};\BC \otimes \cl_{n,n})\rightarrow L_2\left(\Qh;\BC \otimes \cl_{n,n}\right),$$ whose inverse $(\mathcal{F}_{h,\alpha}^{-1} \g)(x,t)=\widehat{\g}_{h,\alpha}(x,t)
$ is given by the Fourier coefficients 
\begin{eqnarray}
\label{FourierInversion}
\widehat{\g}_{h,\alpha}(x,t)=\frac{1}{(2\pi)^{\frac{n}{2}}}\int_{\Qh} (\mathcal{F}_{h,\alpha} \g)(\xi,t) e^{-i x \cdot \xi} d\xi.
\end{eqnarray} 

For the function spaces $\mathcal{S}'(\BR^n_{h,\alpha};\BC\otimes \cl_{n,n})$ and $C^\infty(\Qh;\BC\otimes \cl_{n,n})$, we notice first that $\mathcal{S}'(\BR^n_{h,\alpha};\BC\otimes \cl_{n,n})$, defined as above, is dense in $\ell_2(\BR^n_{h,\alpha};\BC \otimes \cl_{n,n})$. On the other hand, $C^\infty(\Qh;\BC\otimes \cl_{n,n})$ is embedded on the dual space $C^\infty(\Qh;\BC \otimes \cl_{n,n})'$, the so-called space of $\BC\otimes \cl_{n,n}-$valued distributions over $\Qh$ (cf.~\cite[Exercise 3.1.15.]{RuzhanskyT10} \&  \cite[Definition 3.1.25]{RuzhanskyT10}).

As a result, the \textit{Parseval type relation}, involving the sesquilinear forms (\ref{lpInner}) and (\ref{BilinearFormQh}) (cf.~\cite[Definition 3.1.27]{RuzhanskyT10}):
\begin{eqnarray}
\label{Parseval}	\langle \mathcal{F}_{h,\alpha} \f(\cdot,t),\g(\cdot,t
)\rangle_{\Qh}=\left\langle \f(\cdot,t),\widehat{\g}_{h,\alpha}(\cdot,t)\right\rangle_{h,\alpha}
\end{eqnarray}
extends furthermore the isometric isomorphism $$\mathcal{F}_{h,\alpha}:\ell_2(\BR^n_{h,\alpha};\BC \otimes \cl_{n,n})\rightarrow L_2\left(\Qh;\BC \otimes \cl_{n,n}\right)$$ to the mapping $\mathcal{F}_{h,\alpha}:\mathcal{S}'(\BR^n_{h,\alpha};\BC \otimes \cl_{n,n})\rightarrow C^\infty(\Qh;\BC \otimes \cl_{n,n})$.

That allows us to define a \textit{discrete convolution operation} $\star_{h,\alpha}$ between a \textit{discrete distribution} $\f(\cdot,t)$ with membership in $\mathcal{S}'(\BR^n_{h,\alpha};\BC \otimes \cl_{n,n})$, and a \textit{discrete function} $\Phi(x)$ with membership in $\mathcal{S}(\BR^n_{h,\alpha};\BC \otimes \cl_{n,n})$:
\begin{eqnarray}
\label{discreteConvolution} \left(\f(\cdot,t)\star_{h,\alpha} \Phi\right)(x)=\sum_{y\in \BR^n_{h,\alpha}} h^n \Phi(y)\f(y-x,t)
\end{eqnarray}
through the duality condition
\begin{eqnarray*}
	\left\langle~ \f(\cdot,t) \star_{h,\alpha} \Phi,\g(\cdot,t)~\right\rangle_{h,\alpha}=\langle~ \f(\cdot,t),\widetilde{\Phi} \star_{h,\alpha} \g(\cdot,t)~\rangle_{h,\alpha}, & \mbox{with}& \widetilde{\Phi}(x)=[\Phi(-x)]^\dagger.
\end{eqnarray*}

Noteworthy, the \textit{Parseval type relation} (\ref{Parseval}) allows us to show that the following \textit{discrete convolution property} at the level of distributions:
\begin{eqnarray}
\label{ConvolutionFhProperty}\mathcal{F}_{h,\alpha}\left[\f(\cdot,t)\star_{h,\alpha} \Phi\right]=\left(\mathcal{F}_{h,\alpha}\f(\cdot,t)\right)\left(\mathcal{F}_{h,\alpha}\Phi\right)
\end{eqnarray}
 yields straightforwardly from the sequence of identities
\begin{eqnarray*}
	\langle~\mathcal{F}_{h,\alpha}\left[\f(\cdot,t)\star_{h,\alpha} \Phi\right],\g(\cdot,t) ~\rangle_{\Qh}&=&
	\langle~\f(\cdot,t)\star_{h,\alpha} \Phi, \mathcal{F}_{h,\alpha}^{-1}[\g(\cdot,t)]~ \rangle_{h,\alpha}
	\\
	&=&\langle~\f(\cdot,t), \widetilde{\Phi}\star_{h,\alpha}\mathcal{F}_{h,\alpha}^{-1}[\g(\cdot,t)]~ \rangle_{h,\alpha} \\
	&=&\left\langle~\f(\cdot,t),\mathcal{F}_{h,\alpha}^{-1}\left( \mathcal{F}_{h,\alpha}\widetilde{\Phi}~\g(\cdot,t)\right)~ \right\rangle_{h,\alpha} \\
	&=&\left\langle~\mathcal{F}_{h,\alpha}\f(\cdot,t), \mathcal{F}_{h,\alpha}\widetilde{\Phi}~\g(\cdot,t)~ \right\rangle_{h,\alpha} \\
	&=&\langle~\left(\mathcal{F}_{h,\alpha}\f(\cdot,t)\right)\left(\mathcal{F}_{h,\alpha}\Phi\right),\g(\cdot,t) ~\rangle_{\Qh}.
\end{eqnarray*}

\subsection{Discrete Dirac and Discrete Laplacian}\label{discreteOpSub}

Consider now, for each $h>0$, the discrete Laplacian on the lattice $h\BZ^n\subseteq \BR_{h,\alpha}^n$
\begin{eqnarray}
\label{discreteLaplacian}
\displaystyle \Delta_h \f(x,t)=\sum_{j=1}^n
\frac{\f(x+h\e_j,t)+\f(x-h\e_j,t)-2\f(x,t)}{h^2}.
\end{eqnarray}
and the Fourier multiplier of $\mathcal{F}_{h,\alpha}\circ (-\Delta_h)\circ\mathcal{F}_{h,\alpha}^{-1}$
\begin{eqnarray}
\label{SymbolDiscreteLaplace}\displaystyle d_h(\xi)^2=\frac{4}{h^2}\sum_{j=1} \sin^{2}\left(\frac{h\xi_j}{2}\right).
\end{eqnarray} 

By means of the mapping properties associated to the discrete Fourier transform (\ref{discreteFh})
together with the Dirac-K\"ahler discretization on the lattice $\varepsilon\BZ^n$, already considered in \cite{FaustinoMMAS17} (see also \cite[section 1.2]{FaustinoRelativistic18}):
\begin{eqnarray}
\label{DiracEqh}
\begin{array}{lll}
D_\varepsilon \f(x,t)&=&\displaystyle \sum_{j=1}^n\e_j\frac{\f(x+\varepsilon \e_j,t)-\f(x-\varepsilon\e_j,t)}{2\varepsilon}+ \\
&+&	\displaystyle \sum_{j=1}^n\e_{n+j}\frac{2\f(x,t)-\f(x+\varepsilon \e_j,t)-\f(x-\varepsilon \e_j,t)}{2\varepsilon}
\end{array}
\end{eqnarray}
one can further exploit the framework considered in a series of author's previous papers \cite{Faustino16,FaustinoMMAS17} (see also \cite{FaustinoKS07,RSKS10,BRS12,CKKS14} for further comparisons) toward pseudo-differential calculus.
Concretely speaking, the discrete Dirac type operators $$D_{h,\alpha}:\mathcal{S}(\BR^n_{h,\alpha};\BC \otimes \cl_{n,n})\rightarrow \mathcal{S}(\BR^n_{h,\alpha};\BC \otimes \cl_{n,n})$$ satisfying the factorization property $(D_{h,\alpha})^2=-\Delta_h$ can be straightforwardly determined by 
the Fourier multiplier of $\mathcal{F}_{h,\alpha}\circ D_{h,\alpha}\circ\mathcal{F}_{h,\alpha}^{-1}$, defined componentwise by the Clifford-vector-valued function (cf.~\cite[Subsection 2.2.]{FaustinoRelativistic18})
\begin{eqnarray}
\label{DiracSymbol}	
\begin{array}{ccc}
\textbf{z}_{h,\alpha}(\xi)
&=&\displaystyle \sum_{j=1}^n -i\e_j\dfrac{\sin((1-\alpha) h\xi_j)+\sin(\alpha h\xi_j)}{h}+\\
&+&\displaystyle \sum_{j=1}^n \e_{n+j}\dfrac{\cos(\alpha h\xi_j)-\cos((1-\alpha) h\xi_j)}{h}.
\end{array}
\end{eqnarray}

The key property besides this approach is the square condition $\textbf{z}_{h,\alpha}(\xi)^2=d_h(\xi)^2$ between the Fourier multipliers of $\mathcal{F}_{h,\alpha}\circ D_{h,\alpha}\circ\mathcal{F}_{h,\alpha}^{-1}$ and $\mathcal{F}_{h,\alpha}\circ (-\Delta_h)\circ\mathcal{F}_{h,\alpha}^{-1}$, respectively.

As it depicted in \cite[Remark 2.1]{FaustinoRelativistic18}, the limit case $\alpha\rightarrow 0$ 
\begin{eqnarray*}
	\label{DiracSymbol0}	
	\begin{array}{ccc}
		\textbf{z}_{h,0}(\xi)
		&=&\displaystyle \sum_{j=1}^n -i\e_j\dfrac{\sin(h\xi_j)}{h}+\displaystyle \sum_{j=1}^n \e_{n+j}\dfrac{1-\cos(h\xi_j)}{h}.
	\end{array}
\end{eqnarray*}
yields the Fourier multiplier of $\mathcal{F}_{h,\alpha}\circ D_{h}\circ\mathcal{F}_{h,\alpha}^{-1}$ on $h\BZ^n$, encoded by 
the Dirac-K\"ahler discretization $D_h$ (set $\varepsilon=h$ on eq. (\ref{DiracEqh})), whereas in the limit $\alpha\rightarrow\frac{1}{2}$, the symbol $${\bf z}_{h,\frac{1}{2}}(\xi)=\displaystyle \sum_{j=1}^n -i\e_j\dfrac{2\sin(\frac{h\xi_j}{2})}{h}$$ stands for the Fourier multiplier of the self-adjoint discretization of the Dirac operator on the lattice $\frac{h}{2}\BZ^n$.

Considering now the formal $\dagger-$conjugation of (\ref{DiracEqh}) induced by the Clifford algebraic property (\ref{dagconjugation}):
\begin{eqnarray*}
	\label{DiracEqhAdjoint}
	\begin{array}{lll}
		D_\varepsilon^\dagger \f(x,t)&=&\displaystyle \sum_{j=1}^n -\e_j\frac{\f(x+\varepsilon \e_j,t)-\f(x-\varepsilon\e_j,t)}{2\varepsilon}+ \\
		&+&\displaystyle \sum_{j=1}^n \e_{n+j}\frac{2\f(x,t)-\f(x+\varepsilon \e_j,t)-\f(x-\varepsilon \e_j,t)}{2\varepsilon}.
	\end{array}
\end{eqnarray*}
it readily follows from a direct application of the properties of the \textit{discrete Fourier transform} (\ref{discreteFh}) that the operator $D_{h,\alpha}:\mathcal{S}(\BR^n_{h,\alpha};\BC \otimes \cl_{n,n})\rightarrow \mathcal{S}(\BR^n_{h,\alpha};\BC \otimes \cl_{n,n})$ satisfying $$\mathcal{F}_{h,\alpha}(D_{h,\alpha}\g)(\xi)={\bf z}_{h,\alpha}(\xi)~(\mathcal{F}_{h,\alpha}\g)(\xi)$$ is uniquely determined by
\begin{eqnarray}
\label{FractionalDiffDirac} D_{h,\alpha}:=(1-\alpha)D_{(1-\alpha)h}-\alpha D_{\alpha h}^\dagger.
\end{eqnarray}

\section{Time-Changed Dirac-Fokker-Planck equation}\label{TimeChangedFPDSection}

\subsection{The model problem explained}\label{ModelProblemSub}

In this work we propose a time-changed variant of the Dirac equation $\partial_t \Phi(x,t)=i	\mu D_{h,\alpha}\Phi(x,t)$ on $\BR_{h,\alpha}^n \times [0,\infty)$ by adding an extra time-changed perturbation of the order of cut-off that seamlessly describes a discrete counterpart of a \textit{fractional Wiener process} carrying the time-changed diffusion term $Ht^{2H-1}\sigma^2$. 

%%%% NEW PART EXPLAINING THE DERIVATION OF THE MODEL PROBLEM @nelson
In order to proceed, let us entertain with the stochastic process $\{Z_t\}_{t\geq 0}$ defined uniquely via the following stochastic differential equation (SDE) over $\BR_{h,\alpha}^n \times [0,\infty)$:
\begin{eqnarray}
\label{SDEBH}Z_{t}(x)=Z_0(x)+\int_{0}^t -i\mu D_{h,\alpha} Z_s(x)ds+\int_{0}^tdB_{s}^H(x).
\end{eqnarray}

Here $\{B_t^H\}_{t\geq 0}$ stands for the $n-$dimensional \textit{fractional Brownian motion} (fBM) with zero mean ($\mathbb{E}(B_t^H)=0$) and covariance 
\begin{eqnarray*}
\mbox{cov}(B_s^H,B_t^H)&:=&\mathbb{E}(B_s^HB_t^H)-\mathbb{E}(B_s^H)\mathbb{E}(B_t^H)\\
&=&\frac{1}{2}(s^{2H}+t^{2H}-|s-t|^{2H}).
\end{eqnarray*}

Also, we shall assume that the random variable $Z_0$ is independent of $B_t^H$, that is $\mbox{cov}(Z_0,B_t^H)=0$ to ensure the independence of the processes $\{-i\mu D_{h,\alpha}Z_t\}_{t\geq 0}$ and $\{B_t^H\}_{t\geq 0}$, respectively.

For the deduction of the time-changed Dirac-Fokker-Planck (DFP) type equation on the space-time lattice $(x,t)\in \BR_{h,\alpha}^n \times [0,\infty)$, one has to consider the sesquilinear form (\ref{lpInner}) over the {\it right Hilbert module} $\ell_2(\BR^n_{h,\alpha};\BC\otimes \cl_{n,n})$, to represent (\ref{SDEBH}) in its {\it weak} form. Namely, by setting
\begin{eqnarray*}
	dB_s^H(x)=\frac{\sigma^2}{2}\Delta_h Z_s(x)ds^{2H}, & \mbox{with} & s^{2H}=\mbox{cov}(B_s^H,B_s^H)
\end{eqnarray*}
one gets that
\begin{eqnarray}
\label{SDEBHweak}
\begin{array}{lll}
\langle \Phi(x,t),Z_t(x)\rangle_{h,\alpha}&=&\langle \Phi(x,t),Z_0(x)\rangle_{h,\alpha}+\\
&+&\displaystyle \int_{0}^t\left\langle \Phi(x,t),-i\mu D_{h,\alpha}Z_s(x)\right\rangle_{h,\alpha}~ds+\\
&+&\displaystyle \int_{0}^{t}\left\langle \Phi(x,t),\dfrac{\sigma^2}{2}\Delta_hZ_s(x)\right\rangle_{h,\alpha}ds^{2H}.
\end{array}
\end{eqnarray}

Henceforth, one can use the identity $ds^{2H}=2Hs^{2H-1}ds$ and fact that the discrete Dirac operator $D_{h,\alpha}$ (see eq.~(\ref{FractionalDiffDirac})) and the discrete Laplacian $\Delta_h=-(D_{h,\alpha})^2$  (see eq. (\ref{discreteLaplacian})) are self-adjoint operators w.r.t the sesquilinear form (\ref{lpInner}) (cf.~\cite[p.~449]{FaustinoRelativistic18}):
\begin{eqnarray*}
\langle \Phi(x,t),D_{h,\alpha} Z_s(x)\rangle_{h,\alpha}&=&
\langle D_{h,\alpha}\Phi(x,t),Z_s(x)\rangle_{h,\alpha} \\
\langle \Phi(x,t),\Delta_h Z_s(x)\rangle_{h,\alpha}&=&
\langle \Delta_h\Phi(x,t),Z_s(x)\rangle_{h,\alpha} 
\end{eqnarray*}
to derive the following equivalent formulation of (\ref{SDEBHweak}): 
\begin{eqnarray}
\label{SDEBHweak2}
\begin{array}{lll}
\langle \Phi(x,t),Z_t(x)\rangle_{h,\alpha}=\langle \Phi(x,t),Z_0(x)\rangle_{h,\alpha}+\\+\displaystyle \int_{0}^t\langle i\mu D_{h,\alpha} \Phi(x,t)+H\sigma^2s^{2H-1}\Delta_h\Phi(x,t),Z_s(x)\rangle_{h,\alpha}~ ds.
\end{array}
\end{eqnarray}

Moreover, by setting $\Phi(x,0)=\Phi_0(x)$ and letting act the partial derivative $\partial_t$ on both sides of (\ref{SDEBHweak2}),
we obtain that the resulting coupled systems of relations
\begin{eqnarray*}
	\langle \partial_t\Phi(x,t),Z_t(x)\rangle_{h,\alpha}&=&\langle i\mu D_{h,\alpha} \Phi(x,t)+H\sigma^2t^{2H-1}\Delta_h\Phi(x,t),Z_t(x)\rangle_{h,\alpha}
	\\
	\langle \Phi(x,0),Z_t(x)\rangle_{h,\alpha}&=&\langle \Phi_0(x),Z_t(x)\rangle_{h,\alpha}
\end{eqnarray*}
gives rise to the following time-changed Dirac-Fokker-Planck (DFP) type equation on the space-time lattice $(x,t)\in \BR_{h,\alpha}^n \times [0,\infty)$:
\begin{eqnarray}
\label{DiracFokkerPlanck}
\partial_t \Phi(x,t)= i\mu D_{h,\alpha}\Phi(x,t)+\sigma^2Ht^{2H-1}\Delta_h \Phi(x,t),& \Phi(x,0)=\Phi_0(x).
\end{eqnarray}

The above model problem approximates the \textit{discrete massless Dirac equation} in the limit $H\rightarrow 0$. Its right hand side depending on the time variable term $Ht^{2H-1}$ carrying the Hurst parameter $0<H<1$ (see e.g. \cite{HKU11,HRKU11} and the references given there) reflects the time-changed dependence of the discrete Laplacian $\Delta_h$ for values $H\neq \frac{1}{2}$ (cf.~\cite{MandelNess68}). Moreover, the limit conditions $\alpha,H \rightarrow 0$ allows us to recover, from a multivector calculus perspective, the \textit{massless Dirac equation} considered on Rabin's approach \cite{Rabin82} towards the investigation of the \textit{lattice fermion doubling} phenomena (see also \cite[Subsection 3.2]{FaustinoRelativistic18}, on which the solution of the discrete Dirac equation was investigated by means of operational techniques).

Both of this clues allow us to make a reasonable guess that the formulation of the model problem (\ref{DiracFokkerPlanck}) may be viewed as a stochastic rescaling of Wilson's formulation \cite{Wilson74} on which the second order term $\sigma^2Ht^{2H-1}\Delta_h \Phi(x,t)$ encoding the variance function $\sigma_H^2(t)=\sigma^2t^{2H}$ (see also \cite[subsection 3.1]{Hairer05}):
$$\displaystyle \frac{1}{2}\frac{d\sigma_H^2(t)}{dt}=H\sigma^2t^{2H-1}$$ acts as Wilson-like term on the momentum space $\Qh\times [0,\infty)$, in case where $0<H\leq \alpha$ ($0<\alpha<\frac{1}{2}$). 

At this stage, one notice here that in case where $H=\frac{1}{2}$, it is commonly to choose $\sigma^2=h$ as the Wilson parameter $r$ -- that is assumed to be on the interval $0<r\leq 1$ (cf.~\cite[p.~178]{MontvayMunster97}). 
Interesting enough, for values of $H$ in the range $0<H\leq \alpha$ one easily recognizes that the time-dependent diffusion term satisfies the limit condition $\displaystyle \lim_{t\rightarrow \infty}\sigma^2Ht^{2H-1}=0$. 

\begin{remark}[Regularization of Wilson's approach]\label{WilsonRemark}
Despite the choice of $H$ that yields a Wilson-like parameter may be independently taken for values of $H$ on the interval $0<H<\frac{1}{2}$ (sub-diffusive case), one has considered the constraint $H\leq \alpha$ to strongly emphasize that our model problem (\ref{DiracFokkerPlanck}) naturally leads, for special choices of $H$ [bounded above by $\alpha$], to a fractional time-dependent regularization of Wilson's seminal approach \cite{Wilson74} in the limit $\alpha \rightarrow\frac{1}{2}$.   

Noteworthy, the replacement of the discrete Dirac operator $$\displaystyle \frac{1}{2}(D_{h/2}^++D_{h/2}^-):=\lim_{\alpha \rightarrow \frac{1}{2}}D_{h,\alpha},$$ endowed by central difference operators on the lattice $\frac{h}{2}\BZ^n$ (cf.~\cite[p.~449]{FaustinoRelativistic18}), by the family of discrete Dirac operators $D_{h,\alpha}$ ($0<\alpha<\frac{1}{2}$) allows us to rid from the evadedness of Nielsen-Ninomiya's no-go result \cite{NN81} stressed on section \ref{IntroductionSection}.~\textsc{Introduction}, since the chiral symmetries as well as the additional fermion doublers underlying to the Clifford algebra of signature $(n,n)$ are not only preserved for values of the Hurst parameter $H$ in the range $0<H\leq \alpha$, but also for $H=\frac{1}{2}$ under the limit condition $\sigma^2\rightarrow 0$.
\end{remark}

\begin{remark}[Toward fBM]\label{fBMRemark}
	Commonly the diffusion term $\sigma^2$ depends on the Hurst parameter $0<H<1$. And due to the spectral density analysis studied in depth by Mandelbrot and Van Ness in \cite[Section 7]{MandelNess68} it remains natural to consider diffusion terms of the form 
	\begin{eqnarray*}
~\sigma^2=\dfrac{\Gamma(2-2H)}{\pi H(2H-1)}\cos(\pi (H-1)),
	\end{eqnarray*}
where $\Gamma$ stands for the Gamma function defined via the Eulerian integral (\ref{GammaInt}).

Herein, the combination of the identity $$\sigma^2=\frac{\Gamma(2-2H)}{\pi H(1-2H)}\sin\left(\pi\left(\frac{1}{2}-H\right)\right)$$ with the inequalities $0<
{\Gamma(2-2H)}< 1
$ \& $0<2\sin\left(\pi\left(\frac{1}{2}-H\right)\right)<\pi\left(1-2H\right)$, for values of $H$ in the range $0<H<\frac{1}{2}$,
 result into the estimate
\begin{eqnarray*}
	0<H\sigma^2 \left(\frac{1}{h}\right)^{2H-1} <\frac{1}{2}, &\mbox{in the limit} & h\rightarrow 0.
\end{eqnarray*} 

Hence, by assuming in the sub-diffusive case that the Hurst parameter $H$  can be neatly approximated by the fractional parameter $\alpha$ ($0<\alpha<\frac{1}{2}$), carrying the lattice $\BR^n_{h,\alpha}$, the asymptotic condition $h^{1-2\alpha}\approx h^{2-2\alpha}$ in the limit $h\rightarrow 0$ suggests us to choose $$2H\sigma^2\left(\frac{1}{h}\right)^{2H-1}\approx \dfrac{\Gamma(2-2\alpha)}{\pi \left(\frac{1}{2}-\alpha\right)}\sin\left(\pi\left(\frac{1}{2}-\alpha\right)\right)h^{2-2\alpha}$$ as a regularization of the Wilson-like parameter $r=h$ highlighted in \cite[p.~178]{MontvayMunster97} (case $\alpha\rightarrow\frac{1}{2}$) for our model problem (\ref{DiracFokkerPlanck}) in the sub-diffusive case ($0<H<\frac{1}{2}$). Interesting enough, such choice is naturally associated to the Fourier multipliers generated from one-side stable distributions of L\'evy type (see for example \cite[Chapter 3]{MeerSik11}).
\end{remark}

\subsection{A stochastic interpretation of the DFP equation}\label{StochasticSubsection}

In order to provide a stochastic meaning to the solutions of (\ref{DiracFokkerPlanck}), we must impose the normalization condition 
\begin{eqnarray}
\label{quasiProbabilityCondition}	\sum_{x\in \BR_{h,\alpha}^n}h^n\Phi(x,t)=1, & \mbox{for all} & t\geq 0.
\end{eqnarray} 

In the zero-drift case ($\mu=0$) the analysis may be considerably simplified to the study of the discrete fundamental solution of the discrete heat equation (cf.~\cite{BaaskeBRS14}). In particular, if $\Phi(x,0)=\Phi_0(x)$ equals to the \textit{discrete delta distribution} 
\begin{eqnarray*}
	\label{discreteDeltah} \delta_h(x)=\left\{\begin{array}{lll} 
		\dfrac{1}{h^n} & \mbox{if} & x=(0,0,\ldots,0) 
		\\ \ \\
		0 & \mbox{if} & x\neq (0,0,\ldots,0)
	\end{array}\right.,
\end{eqnarray*} the function $\Phi(x,t):=\exp(t\Delta_h)\delta_h(x)$ turns out to be fundamental solution of the \textit{semi-discrete heat operator} $\partial_t-\Delta_h$ (cf.~\cite[Section 4]{BaaskeBRS14}), approximates the semimartingale case in the limit $h\rightarrow 0$ 
($\sigma^2=2$ and $H=\frac{1}{2}$). Interestingly enough, the following representation formula
\begin{eqnarray}
\label{ClosedFormulaBessel} \exp(t\Delta_h)\delta_h(x)&=&\frac{(2\pi)^{\frac{n}{2}}}{h^{n}}e^{-\frac{2nt}{h^2}}  I_{\frac{x_1}{h}}\left(\frac{2t}{h^2}\right)I_{\frac{x_2}{h}}\left(\frac{2t}{h^2}\right) \ldots I_{\frac{x_n}{h}}\left(\frac{2t}{h^2}\right),
\end{eqnarray}
written in terms of the \textit{modified Bessel functions of the first kind} $I_k(u)$ (cf.~\cite[subsection 4.2]{FaustinoRelativistic18}) seamlessy describes to a $n-$ary product of transition probability densities 
carrying a finite sequence of Bessel processes $R^{(\nu_1)}_r,R^{(\nu_2)}_r,\ldots,R^{(\nu_n)}_r$ of order $\nu_j=\frac{|x_j|}{h}\in \mathbb{N}_0$ ($j=1,2,\ldots,n$) (cf.~\cite[pp.~71-76]{BorodinSalm12}).

Indeed, using the fact that the \textit{modified Bessel function} $I_k(r)$ of order $k$ admits the Laplace identity
$$
I_k(r)=\int_{0}^{\infty}\exp\left(-\frac{1}{2}k^2p\right)\theta_r(p)dp
$$
in terms of the (unnormalized) Hartman-Watson distribution 
$$\theta_r(p)=
\frac{r}{\sqrt{2\pi^3u}}\int_{0}^{\infty}\exp\left(-\frac{\pi^2-p^2}{2u}-r\cosh(p)\right)\sinh(p)\sin\left(\frac{\pi p}{u}\right)dp
$$ 
deduced by Marc Yor \cite{Yor80} (see also~\cite[p.~79]{BorodinSalm12}) gives rise to 
\begin{eqnarray}
\label{ClosedFormulaHartmanWatson} 
\begin{array}{lll}
\displaystyle \exp(t\Delta_h)\delta_h(x)=\\  =\displaystyle \frac{(2\pi)^{\frac{n}{2}}}{h^{n}}e^{-\frac{2nt}{h^2}}\displaystyle \int_{[0,\infty)^n} \exp\left(-\sum_{j=1}^n\frac{x_j^2\xi_j}{2h^2}\right)\Theta\left(\xi;\frac{2t}{h^2}\right)d\xi,
\end{array}
\end{eqnarray}
where $d\xi$ denotes the Lebesgue measure and the $n-$ary product
$$ \Theta\left(\xi;\frac{2t}{h^2}\right)=\displaystyle \theta_{\frac{2t}{h^2}}(\xi_1)~\theta_{\frac{2t}{h^2}}(\xi_2)~\ldots ~\theta_{\frac{2t}{h^2}}(\xi_n).$$

With formula (\ref{ClosedFormulaHartmanWatson}) we have a precise probabilistic interpretation for the fundamental solution of $\partial_t-\Delta_h$.
More generally, by taking into account only the normalization condition (\ref{quasiProbabilityCondition}) -- that is, the condition $\Phi(x,t)\geq 0$ will be evaded {\it a priori} -- any Clifford-vector-valued function $\Phi_0(x)$ may be chosen as a \textit{discrete quasi-probability distribution} 
\begin{eqnarray}
\label{discreteQuasiProbability}\mbox{Pr}\left(\displaystyle \sum_{j=1}^n\e_j X_j=x\right)=h^n\Phi_0(x)
\end{eqnarray} carrying a set of \textit{independent and identically distributed} (i.i.d) random variables $X_1,X_2,\ldots,X_n$, with the aim of provide a Bayesian probability meaning to the discrete convolution representation
$$\exp(t\Delta_h)\Phi_0(x)=\sum_{y\in \BR_{h,\alpha}^n}h^n\Phi_0(y)\exp(t\Delta_h)\delta_h(y-x)~$$
of the solution of the discrete heat equation studied in detail in refs. \cite{BaaskeBRS14,CiaurriGRTV17} (see, in particular, \cite[Section 6.]{BaaskeBRS14} and \cite[Section 2]{CiaurriGRTV17}). 

In the zero-diffusion case $(\sigma^2=0)$ some interesting choices for the \textit{likelihood function} $h^n\Phi_0(x)$ are e.g. the Poisson and Mittag-Leffler distributions depicted in \cite[subsection 4.1 \& subsection 4.2]{FaustinoMMAS17}. 

We will come back afterwards to the precise stochastic interpretation for the solutions of the DFP equation (\ref{DiracFokkerPlanck}), based on the proof of {\bf Theorem \ref{mainResultDFP}} in subsection \ref{DFPKleinGordonsub}.

\section{Main Results}\label{MainSection}

\subsection{Semi-discrete Klein-Gordon equations}

In this subsection we focus on the study of solutions associated to the time-changed DFP type equation (\ref{DiracFokkerPlanck}) on the space-time lattice $(x,t)\in \BR^n_{h,\alpha} \times [0,\infty)$
and on the solutions of the semi-discrete Klein-Gordon type equation on $(x,t)\in \BR^n_{h,\alpha} \times [0,\infty)$:
\begin{eqnarray}
\label{KleinGordonParametrix}
\left\{\begin{array}{lll}
\partial_t^2 \varPsi(x,t|p)+4pt\partial_t\varPsi(x,t|p)+\\ \ \\+(2p+4p^2t^2)\varPsi(x,t|p)=\mu^2\Delta_h\varPsi(x,t|p)\\ \ \\
\varPsi(x,0|p)=\Phi_0(x)\\ \ \\ \left[\partial_t \varPsi(x,t|p)\right]_{t=0}=i\mu D_{h,\alpha}\Phi_0(x).
\end{array}\right.
\end{eqnarray}

Before proceeding with the main results, we just want to underscore that the exponentiation operator $\exp\left(i\mu t D_{h,\alpha}\right)$ may be used to generate the solutions of the Klein-Gordon type equation on the lattice (cf.~\cite[Corollary 3.1]{FaustinoRelativistic18}). The following lemma, that will be useful on this subsection and elsewhere, mimics the proof of \cite[Theorem 1.1]{FaustinoRelativistic18}:
\begin{lemma}\label{expitzhLemma}
	Let ${\bf z}_{h,\alpha}(\xi)$ and $d_h(\xi)^2$ be the Fourier multipliers of $\mathcal{F}_{h,\alpha}\circ D_{h,\alpha}\circ \mathcal{F}_{h,\alpha}^{-1}$ and $\mathcal{F}_{h,\alpha}\circ (-\Delta_h)\circ \mathcal{F}_{h,\alpha}^{-1}$, respectively.
	Then, the exponentiation function $\exp(i\mu t{\bf z}_{h,\alpha}(\xi))$ admits the following representation
	\begin{eqnarray*}
		\exp(i\mu t{\bf z}_{h,\alpha}(\xi))&=&\cos\left(\mu t\sqrt{d_h(\xi)^2}\right)+\dfrac{\sin(\mu t\sqrt{d_h(\xi)^2})}{\sqrt{d_h(\xi)^2}}~i{\bf z}_{h,\alpha}(\xi).
	\end{eqnarray*}
\end{lemma}

\proof
First, recall that $\displaystyle \exp\left(i\mu t{\bf z}_{h,\alpha}(\xi)\right)=\cosh\left(i\mu t{\bf z}_{h,\alpha}(\xi)\right)+\sinh\left(i\mu t{\bf z}_{h,\alpha}(\xi)\right)$, whereby
\begin{eqnarray}
\label{SinhCoshitzh}
\begin{array}{lll}
\cosh\left(i\mu t{\bf z}_{h,\alpha}(\xi)\right)&=&\displaystyle \sum_{k=0}^\infty \frac{(\mu t)^{2k}}{(2k)!}\left(i{\bf z}_{h,\alpha}(\xi)\right)^{2k}\\ \ \\ \sinh\left(i\mu t{\bf z}_{h,\alpha}(\xi)\right)&=&\displaystyle\sum_{k=0}^\infty \frac{(\mu t)^{2k+1}}{(2k+1)!}\left(i{\bf z}_{h,\alpha}(\xi)\right)^{2k+1}
\end{array}
\end{eqnarray}
denotes the even resp. odd part of the formal series expansion of $\exp\left(i\mu t{\bf z}_{h,\alpha}(\xi)\right)$.  

From the factorization property ${\bf z}_{h,\alpha}(\xi)^2=d_h(\xi)^2$ we thereby obtain that 
\begin{eqnarray*}
	(i{\bf z}_{h,\alpha}(\xi))^{2k}=&i^{2k}({\bf z}_{h,\alpha}(\xi)^2)^k=&(-1)^k(\sqrt{d_h(\xi)^2})^{2k} \\
	(i{\bf z}_{h,\alpha}(\xi))^{2k+1}=&i{\bf z}_{h,\alpha}(\xi)(i{\bf z}_{h,\alpha}(\xi))^{2k}=&(-1)^k\dfrac{(\sqrt{d_h(\xi)^2})^{2k+1}}{\sqrt{d_h(\xi)^2}}i{\bf z}_{h,\alpha}(\xi),
\end{eqnarray*}
hold for every $k\in \mathbb{N}_0$.

Finally, by substituting the previous relations on the right hand side of (\ref{SinhCoshitzh}), one readily obtain by linearity arguments the following identities, involving the formal series expansions of sine and cosine functions, respectively:
\begin{eqnarray*}
	\cosh\left(i\mu t{\bf z}_{h,\alpha}(\xi)\right)&=&\displaystyle \sum_{k=0}^\infty \frac{(-1)^k(\mu t)^{2k}}{(2k)!}(\sqrt{d_h(\xi)^2})^{2k} \\ &=&\cos\left(\mu t\sqrt{d_h(\xi)^2}\right) \\ \ \\
	\sinh\left(i\mu t{\bf z}_{h,\alpha}(\xi)\right)&=&\displaystyle\sum_{k=0}^\infty \frac{(-1)^k(\mu t)^{2k+1}}{(2k+1)!}\dfrac{(\sqrt{d_h(\xi)^2})^{2k+1}}{\sqrt{d_h(\xi)^2} }i{\bf z}_{h,\alpha}(\xi)\\&=&~\dfrac{\sin\left(\mu t\sqrt{d_h(\xi)^2}\right)}{\sqrt{d_h(\xi)^2}}~i{\bf z}_{h,\alpha}(\xi),
\end{eqnarray*}
completeting the proof of {\bf Lemma \ref{expitzhLemma}}. 
\qed

With the construction furnished in {\bf Lemma \ref{expitzhLemma}} we are able to prove that the exponentiation function $\exp\left(i\mu tD_{h,\alpha}\right)$ generates a solution for the semi-discrete Klein Gordon equation (\ref{KleinGordonParametrix}). That corresponds to the following theorem: 
\begin{theorem}\label{KleinGordonAnsatz}
	For a given Clifford-valued function function $\Phi_0$ with membership in $\mathcal{S}(\BR_{h,\alpha}^n;\BC \otimes \cl_{n,n})$, the ansatz function $$\varPsi(x,t|p)=e^{-pt^2}\left(\cos(\mu t\sqrt{-\Delta_h})\Phi_0(x)+\dfrac{\sin(\mu t\sqrt{-\Delta_h})}{\sqrt{-\Delta_h}}iD_{h,\alpha}\Phi_0(x)\right)$$ 
	satisfies the conditions of the evolution problem (\ref{KleinGordonParametrix}).
\end{theorem}

\proof
First, let us take the ansatz function $$\varPsi(x,t)=\cos(\mu t\sqrt{-\Delta_h})\Phi_0(x)+\dfrac{\sin(\mu t\sqrt{-\Delta_h})}{\sqrt{-\Delta_h}}iD_{h,\alpha}\Phi_0(x),$$ 

By applying the discrete Fourier transform $\mathcal{F}_{h,\alpha}$ on both sides we thereby obtain from {\bf Lemma \ref{expitzhLemma}} that
$$
\left(\mathcal{F}_{h,\alpha}\varPsi(\cdot,t)\right)(\xi)=\cos\left(\mu t\sqrt{d_h(\xi)^2}\right)\left(\mathcal{F}_{h,\alpha}\Phi_0\right)(\xi)+\dfrac{\sin(\mu t\sqrt{d_h(\xi)^2})}{\sqrt{d_h(\xi)^2}}i{\bf z}_{h,\alpha}(\xi)\left(\mathcal{F}_{h,\alpha}\Phi_0\right)(\xi)
$$
corresponds to the representation of $\varPsi(x,t)$ on the momentum space $\Qh\times [0,\infty)$.
A simple computation moreover shows that $\left(\mathcal{F}_h\varPsi(\cdot,t)\right)(\xi)$ provides us a solution for the Cauchy problem on $\Qh\times [0,\infty):$
\begin{eqnarray}
\label{KleinGordonMomentum}
\left\{\begin{array}{lll}
\partial_t^2 \left(\mathcal{F}_{h,\alpha}\varPsi(\cdot,t)\right)(\xi)=-\mu^2d_h(\xi)^2\left(\mathcal{F}_{h,\alpha}\varPsi(\cdot,t)\right)(\xi)\\ \ \\
\left(\mathcal{F}_{h,\alpha}\varPsi(\cdot,0)\right)(\xi)=\left(\mathcal{F}_{h,\alpha}\Phi_0\right)(\xi) \\ \ \\
\left[\partial_t \left(\mathcal{F}_{h,\alpha}\varPsi(\cdot,0)\right)(\xi)\right]_{t=0}=i\mu{\bf z}_{h,\alpha}(\xi)\left(\mathcal{F}_{h,\alpha}\Phi_0\right)(\xi).
\end{array}\right.
\end{eqnarray}

Next, let us take the substitution $\left(\mathcal{F}_{h,\alpha}\varPsi(\cdot,t)\right)(\xi)=e^{pt^2}\left(\mathcal{F}_{h,\alpha}\varPsi(\cdot,t|p)\right)(\xi)$ on (\ref{KleinGordonMomentum}) for a given $p\geq 0$. 

Clearly, one has $$\left(\mathcal{F}_{h,\alpha}\varPsi(\cdot,0)\right)(\xi)=\left(\mathcal{F}_{h,\alpha}\varPsi(\cdot,0;p)\right)(\xi)=\left(\mathcal{F}_{h,\alpha}\Phi_0\right)(\xi).$$ 

On the other hand, a straightforward computation based on the Leibniz rule moreover shows that
\begin{eqnarray*}
	\left[\partial_t\left(e^{pt^2}\left(\mathcal{F}_{h,\alpha}\varPsi(\cdot,t|p)\right)(\xi)~ \right)\right]_{t=0}&=&\left[e^{pt^2}\left(~\partial_t\left(\mathcal{F}_{h,\alpha}\varPsi(\cdot,t|p)\right)(\xi)+2pt\left(\mathcal{F}_{h,\alpha}\varPsi(\cdot,t|p)\right)(\xi)~\right)\right]_{t=0} \\
	&=&i{\bf z}_{h,\alpha}(\xi)\left(\mathcal{F}_{h,\alpha}\Phi_0\right)(\xi) \\ \ \\
	\partial_t^2\left(e^{pt^2}\left(\mathcal{F}_{h,\alpha}\varPsi(\cdot,t|p)\right)(\xi)\right)&=&\partial_t\left[e^{pt^2}\left(\partial_t\left(~\mathcal{F}_{h,\alpha}\varPsi(\cdot,t|p)\right)(\xi)+2pt\left(\mathcal{F}_{h,\alpha}\varPsi(\cdot,t|p)\right)(\xi)~\right)\right] \\
	&=&e^{pt^2}\left[(\partial_t+2pt)\left(\partial_t\left(\mathcal{F}_{h,\alpha}\varPsi(\cdot,t|p)\right)(\xi)+2pt\left(\mathcal{F}_{h,\alpha}\varPsi(\cdot,t|p)\right)(\xi)\right)\right] \\
	&=&e^{pt^2}\left(\partial_t^2+4pt\partial_t+2p+4p^2t^2\right)\left(\mathcal{F}_{h,\alpha}\varPsi(\cdot,t|p)\right)(\xi).
\end{eqnarray*}

From the above set of relations one can therefore conclude that $\left(\mathcal{F}_h\varPsi(\cdot,t)\right)(\xi)$ is a solution of the semi-discrete Cauchy problem
\begin{eqnarray}
\label{KleinGordonParametrixMomentum}
\left\{\begin{array}{lll}
\partial_t^2 \left(\mathcal{F}_{h,\alpha}\varPsi(\cdot,t|p)\right)(\xi)+4pt\partial_t\left(\mathcal{F}_{h,\alpha}\varPsi(\cdot,t|p)\right)(\xi)+\\ \ \\+(2p+4p^2t^2)\left(\mathcal{F}_{h,\alpha}\varPsi(\cdot,t|p)\right)(\xi)
=-\mu^2d_h(\xi)^2\left(\mathcal{F}_{h,\alpha}\varPsi(\cdot,t|p)\right)(\xi)\\ \ \\
\left(\mathcal{F}_{h,\alpha}\varPsi(\cdot,0;p)\right)(\xi)=\left(\mathcal{F}_{h,\alpha}\Phi_0\right)(\xi) \\ \ \\
\left[\partial_t\left(\mathcal{F}_{h,\alpha}\varPsi(\cdot,t|p)\right)(\xi)~\right]_{t=0}=i\mu{\bf z}_{h,\alpha}(\xi)\left(\mathcal{F}_{h,\alpha}\Phi_0\right)(\xi).
\end{array}\right.
\end{eqnarray}

Finally, by taking the inverse of the discrete Fourier transform $\mathcal{F}_{h,\alpha}$ on both sides of (\ref{KleinGordonParametrixMomentum}) we conclude that
$$\varPsi(x,t|p)=e^{-pt^2}\left(\cos(\mu t\sqrt{-\Delta_h})\Phi_0(x)+\dfrac{\sin(\mu t\sqrt{-\Delta_h})}{\sqrt{-\Delta_h}}~iD_{h,\alpha}\Phi_0(x)\right)$$ 
is a solution of (\ref{KleinGordonParametrix}).
\qed

\subsection{Time-Changed DFP vs. Klein-Gordon}\label{DFPKleinGordonsub}

Let us turn again our attention to the time-changed DFP (\ref{DiracFokkerPlanck}) on the space-time lattice $\BR_{h,\alpha}^n\times [0,\infty)$. We notice that on the momentum space $\Qh \times [0,\infty)$, the equation (\ref{DiracFokkerPlanck}) reads as
\begin{eqnarray}
\label{DiracFokkerPlanckMomentum}
\begin{array}{lll}
\partial_t (\mathcal{F}_{h,\alpha}\Phi(\cdot,t))(\xi)= \left(i\mu {\bf z}_{h,\alpha}(\xi)-\sigma^2Ht^{2H-1}d_h(\xi)^2\right) (\mathcal{F}_{h,\alpha}\Phi(\cdot,t))(\xi), \\ \ \\ (\mathcal{F}_{h,\alpha}\Phi(\cdot,0))(\xi)=(\mathcal{F}_{h,\alpha}\Phi_0)(\xi),
\end{array}
\end{eqnarray}
upon the application of the discrete Fourier transform (\ref{discreteFh}). 

Considering now the exponentiation function $\exp\left(i\mu t{\bf z}_{h,\alpha}(\xi)-\frac{\sigma^2t^{2H}}{2}d_h(\xi)^2\right)$, we recall that
\begin{eqnarray}
\label{expitzhdhProduct}
\begin{array}{lll}
\displaystyle \exp\left(i\mu t{\bf z}_{h,\alpha}(\xi)-\frac{\sigma^2t^{2H}}{2}d_h(\xi)^2\right)=\\ 
\displaystyle =\exp\left(-\frac{\sigma^2t^{2H}}{2}d_h(\xi)^2\right)\exp\left(i\mu t{\bf z}_{h,\alpha}(\xi)\right)
\end{array}
\end{eqnarray}
results from the fact that ${\bf z}_{h,\alpha}(\xi)$ commutes with $d_h(\xi)^2$. Thus
\begin{eqnarray}
\label{DFPMomentumSolution}
(\mathcal{F}_{h,\alpha}\Phi(\cdot,t))(\xi)=\exp\left(i\mu t{\bf z}_{h,\alpha}(\xi)-\frac{\sigma^2t^{2H}}{2}d_h(\xi)^2\right)(\mathcal{F}_{h,\alpha}\Phi_0)(\xi)
\end{eqnarray}
corresponds to the representation of the solution of the evolution equation (\ref{DiracFokkerPlanck}) on the momentum space $\Qh\times [0,\infty)$, since $(\mathcal{F}_{h,\alpha}\Phi(\cdot,0))(\xi)=(\mathcal{F}_{h,\alpha}\Phi_0)(\xi)$ and
\begin{eqnarray*}
	\partial_t (\mathcal{F}_{h,\alpha}\Phi(\cdot,t))(\xi)
	&=&\exp\left(-\frac{\sigma^2t^{2H}}{2}d_h(\xi)^2\right)\left[\partial_t\exp\left(i\mu t{\bf z}_{h,\alpha}(\xi)\right)\right](\mathcal{F}_{h,\alpha}\Phi_0)(\xi)+ \\
	&+& \left[\partial_t\exp\left(-\frac{\sigma^2t^{2H}}{2}d_h(\xi)^2\right)\right]\exp\left(i\mu t{\bf z}_{h,\alpha}(\xi)\right)(\mathcal{F}_{h,\alpha}\Phi_0)(\xi)\\
	&=&(i\mu{\bf z}_{h,\alpha}(\xi)-\sigma^2Ht^{2H-1}d_h(\xi)^2)(\mathcal{F}_{h,\alpha}\Phi(\cdot,t))(\xi).
\end{eqnarray*} 

So if we take the discrete convolution property (\ref{ConvolutionFhProperty}) underlying to mapping property $\mathcal{F}_{h,\alpha}:\mathcal{S}'(\BR^n_{h,\alpha};\BC \otimes \cl_{n,n})\rightarrow C^\infty(\Qh;\BC \otimes \cl_{n,n})$, we thus have proved the following:

\begin{theorem}\label{mainResultDFP}
	Let $\Phi_0$ be Clifford-valued function membership in $\mathcal{S}(\BR_{h,\alpha}^n;\BC \otimes \cl_{n,n})$, and 
	$\FH$ a kernel function defined by the integral formula 
	\begin{eqnarray*}
		\FH(x,t|\mu,\sigma^2)=\displaystyle \frac{1}{(2\pi)^{\frac{n}{2}}} \int_{\Qh} \exp\left(-\frac{\sigma^2t^{2H}}{2}d_h(\xi)^2\right)\exp\left(i\mu t{\bf z}_{h,\alpha}(\xi)\right) e^{-i x\cdot \xi}~d\xi .
	\end{eqnarray*}
	
	Then we have the following:
	\begin{enumerate}
		\item[{\bf (i)}] The ansatz 		\begin{eqnarray}
		\label{ansatzFPD} \Phi(x,t)=\exp\left(i\mu tD_{h,\alpha}+\frac{\sigma^2t^{2H}}{2}\Delta_h\right)\Phi_0(x)
		\end{eqnarray} solves the Dirac-Fokker-Planck equation (\ref{DiracFokkerPlanck}) on the space-time lattice $\BR^n_{h,\alpha}\times [0,\infty)$. \label{SemigroupSolutionDFP}
		
		\item[{\bf (ii)}] $\Phi(x,t)$ is uniquely determined by the discrete convolution representation 
		\begin{eqnarray*}
			\label{discreteConvolutionFPD} (\FH(\cdot,t|\mu,\sigma^2)\star_{h,\alpha} \Phi_0)(x)=\sum_{y \in \BR_{h,\alpha}^n}h^n\Phi_0(y) \FH(x-y,t|\mu,\sigma^2).
		\end{eqnarray*} 
	\end{enumerate} 
\end{theorem}

In order to obtain an interplay with the Klein-Gordon equation associated to the Cauchy problem (\ref{KleinGordonParametrix}) we would like to stress first that the product rule (\ref{expitzhdhProduct}) allows also to recast the operational formula (\ref{ansatzFPD}) as
\begin{eqnarray*}
	\Phi(x,t)=\exp\left(\frac{\sigma^2t^{2H}}{2}\Delta_h\right)\varPsi(x,t), &\mbox{with}& \varPsi(x,t)=\exp\left(i\mu tD_{h,\alpha}\right)\Phi_0(x)
\end{eqnarray*}
so that
\begin{eqnarray}
	\label{discreteConvolutionFPD}\Phi(x,t)=\sum_{y \in \BR_{h,\alpha}^n}h^n\varPsi(y,t) \FH(x-y,t|0,\sigma^2)
\end{eqnarray}
corresponds to an equivalent formulation for the convolution representation provided by \textbf{Theorem \ref{mainResultDFP}}. Essentially, that involves the discrete convolution between the solution $\varPsi(x,t|0):=\varPsi(x,t)$ of (\ref{KleinGordonParametrix}) and the kernel function $\FH(x,t|0,\sigma^2)$. 

With $\FH$, described as before, a closed formula for $$\FH(x,t|0,\sigma^2)=\exp\left(\frac{\sigma^2 t^{2H}}{2}\Delta_h\right)\delta_h(x)$$ may be easily obtained upon the replacement $t\rightarrow \frac{\sigma^2t^{2H}}{2}$ on the right hand sides of (\ref{ClosedFormulaBessel}) and (\ref{ClosedFormulaHartmanWatson}) so that $\FH(x-y,t|0,\sigma^2)=e^{-\frac{n\sigma^2t^{2H}}{h^2}} \textbf{N}_H(x-y,t|\sigma^2)$, with
\begin{eqnarray}
\label{ClosedFormulaHartmanWatsonH} 
\begin{array}{lll}
\displaystyle 	\textbf{N}_H(x-y,t|\sigma^2)=\\ \ \\ \displaystyle =\frac{(2\pi)^{\frac{n}{2}}}{h^{n}} I_{\frac{x_1}{h}}\left(\frac{t^{2H}\sigma^2}{h^2}\right)I_{\frac{x_2}{h}}\left(\frac{t^{2H}\sigma^2}{h^2}\right) \ldots I_{\frac{x_n}{h}}\left(\frac{t^{2H}\sigma^2}{h^2}\right)\\ \ \\
=\displaystyle \frac{(2\pi)^{\frac{n}{2}}}{h^{n}}\int_{[0,\infty)^n} \exp\left(-\sum_{j=1}^n\frac{(x_j-y_j)^2\xi_j}{2h^2}\right)\Theta\left(\xi;\frac{t^{2H}\sigma^2}{h^2}\right)d\xi.
\end{array}
\end{eqnarray}

\begin{remark}
	In case that the initial condition $\Phi(x,0)=\Phi_0(x)$ endows the {\it quasi-probability distribution} (\ref{discreteQuasiProbability}), it is straightforward to see from statement {\bf (i)} of {\bf Theorem \ref{mainResultDFP}} that the solution $\Phi(x,t)$ encoded by the stochastic process $\{Z_t\}_{t\geq 0}$ (see subsection \ref{ModelProblemSub}) satisfies the {\it quasi-probability condition} (\ref{quasiProbabilityCondition}) fixed in subsection \ref{StochasticSubsection}.
	
	However the resulting convolution representation for $\Phi(x,t)$ obtained in statement {\bf (ii)} of {\bf Theorem \ref{mainResultDFP}} -- and recasted in eq. (\ref{discreteConvolutionFPD}) in terms of the solution $\varPsi(x,t|0):=\varPsi(x,t)$ of the Klein-Gordon problem (\ref{KleinGordonParametrix}) -- does not allows us to interpret the mapping $x\mapsto h^n\Phi(x,t)$ as {\it likelihood} distribution in the Bayesian sense, even if the mapping $x\mapsto h^n\Phi_0(x)$ defines a {\it discrete probability distribution} satisfying the null condition $D_{h,\alpha}\Phi_0(x)=0$ (an analytic condition in disguise).  
	
	Indeed, in the view of {\bf Theorem \ref{KleinGordonAnsatz}}, the null condition $D_{h,\alpha}\Phi_0(x)=0$ only assures that $$\exp\left(i\mu tD_{h,\alpha}\right)\Phi_0(x)=\cos(\mu t\sqrt{-\Delta_h})\Phi_0(x)$$ 
	is at most a real-valued function, even if $\Phi(x,0)=\Phi_0(x)$ is a real-valued function satisfying the \textit{discrete probability distribution} constraints $h^n\Phi_0(x)\geq 0$ and (\ref{quasiProbabilityCondition}).
\end{remark}

Next, we will make use of the Laplace identity (\ref{LevyDistributionsWright}) involving the L\'evy one-sided distribution $L_\nu(u)=\dfrac{1}{u}{~}_0\Psi_1
\left[\begin{array}{l|}    \\
(0,-\nu) 
\end{array}~ \dfrac{1}{u^\nu} \right]$ (see eq. 
(\ref{LevyDistributions}) of Appendix \ref{GeneralizedWrightSub})
 to relate the solutions of the time-changed Dirac-Fokker-Planck equation (\ref{DiracFokkerPlanck}) with the solutions of semi-discrete Klein-Gordon equation (\ref{KleinGordonParametrix}).

\begin{corollary}\label{LevyRepCorollary}
Let  $\Phi(x,t)$ be the solution of the time-changed DFP equation (\ref{DiracFokkerPlanck}) provided by \textbf{Theorem \ref{mainResultDFP}} and
$\varPsi(x,t|p)$ the solution of the differential-difference Klein-Gordon equation (\ref{KleinGordonParametrix}) provided by \textbf{Theorem \ref{KleinGordonAnsatz}}

Then, we have the following:
\begin{enumerate}

\item On the momentum space $\Qh \times [0,\infty)$, the solutions $\Phi(x,t)$ and $\varPsi(x,t|p)$ are interrelated by the operational representation
\begin{eqnarray}
\label{LevyDistributionDFP}	(\mathcal{F}_{h,\alpha}\Phi(\cdot,t))(\xi)
	=\nonumber \\ =\displaystyle \int_{0}^{\infty} \left(\mathcal{F}_{h,\alpha}\varPsi(\cdot,t|p)\right)(\xi)~{~}_0\Psi_1
	\left[\begin{array}{l|}    \\
	(0,-H) 
	\end{array}~ \frac{\sigma^2}{2p^H}d_h(\xi)^2 \right]\frac{dp}{p},
\end{eqnarray}
with $${~}_0\Psi_1
\left[\begin{array}{l|}    \\
(0,-H) 
\end{array}~ \frac{\sigma^2}{2p^H}d_h(\xi)^2 \right]=\displaystyle p~\left(\frac{\sigma^{2} d_h(\xi)^{2}}{2}\right)^{-\frac{1}{H}} L_{H}\left(p~\left(\frac{\sigma^{2} d_h(\xi)^{2}}{2}\right)^{-\frac{1}{H}}\right).$$
\label{SemigroupSolutionIntegral}
\item On the space-time lattice $\BR_{h,\alpha}^n \times [0,\infty)$, the solutions $\Phi(x,t)$ and $\varPsi(x,t|p)$ are interrelated by the \textit{discrete convolution representation}
$$\Phi(x,t)=\sum_{y \in \BR_{h,\alpha}^n}h^n	\widehat{\varPsi}_H(y,t)	\textbf{N}_H(x-y,t|\sigma^2),$$
with
	\begin{eqnarray}
\label{LaplaceKG}
\begin{array}{lll}

\widehat{\varPsi}_H(y,t)&=&\displaystyle \int_{0}^{\infty}\varPsi(y,t|p) {~}_0\Psi_1
\left[\begin{array}{l|}    \\
(0,-H) 
\end{array}~ \frac{n\sigma^2}{h^2p^H} \right]\frac{dp}{p} \\ \ \\
&=&\displaystyle \int_{0}^{\infty}\varPsi(y,t|p)~\left(\frac{n\sigma^2}{h^2}\right)^{-\frac{1}{H}} L_{H}\left(p\left(\frac{n\sigma^2}{h^2}\right)^{-\frac{1}{H}}\right) dp.
\end{array}
\end{eqnarray}
\label{ConvolutionSolutionDFP} 
\end{enumerate}

\end{corollary}

\proof
For the proof of statement (\ref{SemigroupSolutionIntegral}), we recall that for the substitution $s=\left(\frac{\sigma^2}{2}d_h(\xi)^2\right)^{\frac{1}{H}}t^2$ on both sides of (\ref{LevyDistributionsWright}), the sequence of identities
\begin{eqnarray*}
\label{LaplaceSubordinator}
\begin{array}{lll}
\exp\left(-\dfrac{\sigma^2t^{2H}}{2}d_h(\xi)^2\right)= \\ \ \\  \displaystyle =\int_{0}^{\infty} e^{-u\left(\frac{\sigma^2}{2}d_h(\xi)^2\right)^{\frac{1}{H}}t^2}{~}_0\Psi_1
\left[\begin{array}{l|}    \\
(0,-H) 
\end{array}~ \dfrac{1}{u^H} \right]~\frac{du}{u} \\ \ \\ 
=\displaystyle \int_{0}^{\infty} e^{-pt^2}{~}_0\Psi_1
\left[\begin{array}{l|}    \\
(0,-H) 
\end{array}~ \frac{\sigma^2}{2p^H}d_h(\xi)^2 \right]\frac{dp}{p}
\end{array}
\end{eqnarray*}
yield straightforwardly from the change of variable $p=u\left(\frac{\sigma^2}{2}d_h(\xi)^2\right)^{-\frac{1}{H}}$. 

Thus,
\begin{eqnarray*}
(\mathcal{F}_{h,\alpha}\Phi(\cdot,t))(\xi)
= \exp\left(-\frac{\sigma^2t^{2H}}{2}d_h(\xi)^2\right)\exp\left(i\mu t{\bf z}_{h,\alpha}(\xi)\right)(\mathcal{F}_{h,\alpha}\Phi_0)(\xi) \\
\\= \displaystyle \int_{0}^{\infty} \exp(-pt^2)\exp(i\mu t{\bf z}_{h,\alpha}(\xi))~{~}_0\Psi_1
\left[\begin{array}{l|}    \\
	(0,-H) 
\end{array}~ \frac{\sigma^2}{2p^H}d_h(\xi)^2 \right]\frac{dp}{p}.
\end{eqnarray*}

Now, from the combination of \textbf{Lemma \ref{SinhCoshitzh}} with \textbf{Proposition \ref{KleinGordonAnsatz}} we realize that $\exp(-pt^2)\exp(i\mu t{\bf z}_{h,\alpha}(\xi))(\mathcal{F}_{h,\alpha}\Phi_0)(\xi)$ equals to $\left(\mathcal{F}_{h,\alpha}\varPsi(\cdot,t|p)\right)(\xi)$ so that the previous integral identity becomes then
\begin{eqnarray*}
(\mathcal{F}_{h,\alpha}\Phi(\cdot,t))(\xi)
	&=&\displaystyle \int_{0}^{\infty}\left(\mathcal{F}_{h,\alpha}\varPsi(\cdot,t|p)\right)(\xi)~ {~}_0\Psi_1
	\left[\begin{array}{l|}    \\
		(0,-H) 
	\end{array}~ \frac{\sigma^2}{2p^H}d_h(\xi)^2 \right]\frac{dp}{p}.
\end{eqnarray*}

For the proof of (\ref{ConvolutionSolutionDFP}), notice first that 
\begin{eqnarray*}
\Phi(x,t)&=&\sum_{y\in \BR^n_{h,\alpha}}h^n e^{-\frac{n\sigma^2 t^{2H}}{h^2}}\varPsi(y,t)\textbf{N}_H(x-y,t|\sigma^2).
\end{eqnarray*} 

Then, in the same order of ideas of the proof of statement (\ref{SemigroupSolutionIntegral}), we employ the Laplace identity 
$$e^{-\frac{n\sigma^2 t^{2H}}{h^2}}=\int_{0}^{\infty}e^{-pt^2}{~}_0\Psi_1
\left[\begin{array}{l|}    \\
(0,-H) 
\end{array}~ \frac{n\sigma^2}{h^2p^H} \right]\frac{dp}{p}$$
derived from (\ref{LevyDistributionsWright}) of Appendix \ref{GeneralizedWrightSub} to conclude that $e^{-\frac{n\sigma^2 t^{2H}}{h^2}}\varPsi(y,t)$ equals to $\widehat{\varPsi}_H(y,t)$, as desired.
\qed

\subsection{Solution representation through generalized Wright functions}\label{GeneralizedWrightSubsection}

We have essentially used on the proof of \textbf{Corollary \ref{LevyRepCorollary}} that the solution $\Phi(x,t)$ can be represented as a discrete convolution between the kernel function (\ref{ClosedFormulaHartmanWatsonH}) and the function 
\begin{eqnarray*}
e^{-\frac{n\sigma^2 t^{2H}}{h^2}}\varPsi(y,t)=\displaystyle \frac{1}{(2\pi)^{\frac{n}{2}}} \int_{\Qh} e^{-\frac{n\sigma^2 t^{2H}}{h^2}}\exp\left(i\mu t{\bf z}_{h,\alpha}(\xi)\right)(\mathcal{F}_{h,\alpha}\Phi_0)(\xi) e^{-i y\cdot \xi}~d\xi.
\end{eqnarray*}

Bearing in mind the result obtained in \textbf{Lemma \ref{expitzhLemma}}, we know already from the framework developed in \cite[Section 3]{FaustinoRelativistic18} (see, in particular, \cite[Theorem 3.1.]{FaustinoRelativistic18}) that the function $e^{-\frac{n\sigma^2 t^{2H}}{h^2}}\varPsi(x,t)$ described as above may be reformulated as a discrete convolution 
involving the kernel functions
\begin{eqnarray}
\label{kernelFHbeta}
\begin{array}{ccc}
\displaystyle \KH^{\left(0\right)}(y,t|\mu,\sigma^2)=\frac{1}{(2\pi)^{\frac{n}{2}}} \int_{\Qh} e^{-\frac{n\sigma^2 t^{2H}}{h^2}}\cos(\mu t\sqrt{d_h(\xi)^2}) e^{-i y\cdot \xi}~d\xi \\ \ \\
\displaystyle \KH^{\left(1\right)}(y,t|\mu,\sigma^2)= \frac{1}{(2\pi)^{\frac{n}{2}}} \int_{\Qh} e^{-\frac{n\sigma^2 t^{2H}}{h^2}}\frac{\sin(\mu t\sqrt{d_h(\xi)^2})}{\sqrt{d_h(\xi)^2}} e^{-i y\cdot \xi}~d\xi.
\end{array} 	
\end{eqnarray}

That is,
\begin{eqnarray*}
e^{-\frac{n\sigma^2 t^{2H}}{h^2}}\varPsi(y,t)&=&(\KH^{(0)}(\cdot,t|\mu,\sigma^2)\star_{h,\alpha} \Phi_0)(y)+(\KH^{(1)}(\cdot,t|\mu,\sigma^2)\star_{h,\alpha} iD_{h,\alpha}\Phi_0)(y)\\ \ \\ &=&\sum_{x\in \BR_{h,\alpha}^n}\Phi_0(x)~\KH^{\left(0\right)}(y-x,t|\mu,\sigma^2)+\\
&+&\sum_{x\in \BR_{h,\alpha}^n}i D_{h,\alpha}\Phi_0(x)~\KH^{\left(1\right)}(y-x,t|\mu,\sigma^2).
\end{eqnarray*}

To obtain analytic representations for $\KH^{\left(0\right)}$ and $\KH^{\left(1\right)}$ we are going to derive identities involving the generalized Wright functions with the aid of the Mellin transform (see Appendix \ref{FractionalCalcAppendix}).
Before stating the main construction of this section, first define the auxiliar kernel function ${\bf W}_H^{(\beta)}\left(y,\mu t ~|~\omega,\dfrac{n\sigma^2}{h^2}\right)$ via integral eq. (\ref{WrightIntegralH}):
\begin{eqnarray}
	\label{WrightIntegralH}
%	\begin{array}{lll}
{\bf W}_H^{(\beta)}\left(y,\mu t ~|~\omega,\dfrac{n\sigma^2}{h^2}\right)=\\ \nonumber  = \displaystyle \frac{1}{(2\pi)^{\frac{n}{2}}}\int_{\Qh} {~}_1\Psi_1
\left[\begin{array}{l|} \left(\frac{\beta+\omega}{2H},\frac{1}{H}\right)   \\
\left(\beta+\frac{1}{2},1\right) 
\end{array} -\frac{\mu^2 t^2d_h(\xi)^2}{4}\left(\frac{n\sigma^2}{h^2}\right)^{-\frac{1}{H}}\right]e^{-i y\cdot \xi}d\xi.\nonumber
%	\end{array}
	\end{eqnarray} 

We note that from direct application of \cite[Theorem 1]{KilbasSaigoTrujillo02} (see also subsection \ref{GeneralizedWrightSub} of Appendix \ref{FractionalCalcAppendix})
the series expansion of Wright type ${~}_1\Psi_1$ appearing on the integral (\ref{WrightIntegralH})
is uniformly convergent for values of $H$ in the range $\frac{1}{2}\leq H<1$ so that one can only interchange term-by-term of the series with the integral under such constraint. 
In particular, we note that the aforementioned series expansion:
\begin{itemize}
	\item Is uniformly convergent for all $t\geq 0$ in case of $\frac{1}{2}<H<1$ (yields from the condition $1-\frac{1}{H}>-1$);
	\item For $H=\frac{1}{2}$ we can only assure the uniformly convergence of the series of ${~}_1\Psi_1$ type on the compact interval that yield from the inequality $\left|\lambda\right|\leq \rho$, with $\lambda=-\frac{\mu^2 t^2d_h(\xi)^2}{4}\left(\frac{n\sigma^2}{h^2}\right)^{-\frac{1}{H}}$ \& $\rho=\frac{1}{{\left(\frac{1}{2}\right)}^{\frac{1}{2}}}$, whenever the parameter $\omega$ appearing on ${~}_1\Psi_1$ satisfies the condition $\mbox{Re}\left(\omega\right)<0$ (that yields from the constraint $\mbox{Re}\left(\kappa\right)>\frac{1}{2}$, with $\kappa=(\beta+\frac{1}{2})-(\beta+\omega)$).
\end{itemize}

 Since we are interested on the description of the solutions of the DFP equation (\ref{DiracFokkerPlanck}) the space-time lattice $\BR^n_{h,\alpha}\times [0,\infty)$ depending upon the fractional parameter $0<\alpha<\frac{1}{2}$, that justifies the introduction of the sufficient condition $\alpha+\frac{1}{2}\leq H<1$ on the statement of the following theorem.

\begin{theorem}\label{MellinBarnesTheorem}
	Let 
$\KH^{\left(\beta\right)}$ resp. ${\bf W}_H^{\left(\beta\right)}$ be the kernel functions defined via eq. (\ref{kernelFHbeta}) resp. (\ref{WrightIntegralH}).
 In case where the condition $\alpha+\frac{1}{2}\leq H<1$ is imposed to the Hurst parameter $H$, there holds that ${\bf W}_H^{\left(\beta\right)}$ converges uniformly in the space-time lattice $\BR^n_{h,\alpha}\times [0,\infty)$.
 
Moreover, for $\beta=0,1$, the kernel functions $\KH^{\left(\beta\right)}$ admits the Mellin-Barnes representation formula \begin{eqnarray}
\label{MellinBarnesIntegralH}
\begin{array}{lll}
\KH^{\left(\beta\right)}(y,t|\mu,\sigma^2)=\\ \ \\ =\displaystyle\frac{1}{2\pi i}\int_{c-i\infty}^{c+i\infty}~\frac{\sqrt{\pi}\left(\frac{\mu}{2}\right)^\beta}{H}\left(\frac{n\sigma^2}{h^2}\right)^{-\frac{\beta+\omega}{2H}}~{\bf W}_H^{(\beta)}\left(y,\mu t ~|~\omega,\dfrac{n\sigma^2}{h^2}\right)~t^{-\omega}d\omega. %\nonumber
\end{array}
\end{eqnarray}

\end{theorem}
\proof
From the discussion taken previously, we have seen that the sufficient condition $\alpha+\frac{1}{2}\leq H<1$ assures the uniformly convergence of the auxiliar function ${\bf W}_H^{\left(\beta\right)}$ defined via eq.~(\ref{WrightIntegralH}). Thus, it remains to prove only the closed formula (\ref{MellinBarnesIntegralH}).

Firstly, we recall that in view of \textbf{Lemma \ref{expitzhLemma}} and of  eqs. (\ref{WrighCosine}) and (\ref{WrighSinc}) (see subsection \ref{GeneralizedWrightSub} of Appendix \ref{FractionalCalcAppendix}), one can represent the Fourier multipliers
\begin{eqnarray*}
 \displaystyle e^{-\frac{n\sigma^2 t^{2H}}{h^2}}\cos(\mu t\sqrt{d_h(\xi)^2}) &\mbox{and} &\displaystyle e^{-\frac{n\sigma^2 t^{2H}}{h^2}}\frac{\sin(\mu t\sqrt{d_h(\xi)^2})}{\sqrt{d_h(\xi)^2}}
\end{eqnarray*}
appearing on (\ref{kernelFHbeta}) as

\begin{eqnarray}
\label{WrightExpansionCosPsiH}
\begin{array}{lll}
e^{-\frac{n\sigma^2 t^{2H}}{h^2}}\cos(\mu t\sqrt{d_h(\xi)^2})=\\ \ \\=\sqrt{\pi}~e^{-\frac{n\sigma^2 t^{2H}}{h^2}}{~}_0\Psi_1
\left[\begin{array}{l|}    \\
	\left(\frac{1}{2},1\right) 
\end{array} -\dfrac{\mu^2 t^2d_h(\xi)^2 }{4}\right]\\ \ \\
\end{array}
\\
\label{WrightExpansionSincPsiH}
\begin{array}{lll}
e^{-\frac{n\sigma^2 t^{2H}}{h^2}}\dfrac{\sin(\mu t\sqrt{d_h(\xi)^2})}{\sqrt{d_h(\xi)^2}}=\\ \ \\=\displaystyle \dfrac{\mu t\sqrt{\pi}}{2}e^{-\frac{n\sigma^2 t^{2H}}{h^2}}{~}_0\Psi_1
\left[\begin{array}{l|}    \\
	\left(\frac{3}{2},1\right) 
\end{array} -\frac{\mu^2 t^2d_h(\xi)^2}{4} \right].
\end{array}
\end{eqnarray}

Thus, the computation of the kernel functions (\ref{kernelFHbeta}) may be reformulated by means of the compact formula ($\beta=0,1$)
\begin{eqnarray}
%\begin{array}{lll}
\label{compactFH}\KH^{\left(\beta\right)}(y,t|\mu,\sigma^2)=\\ \nonumber \\  =\displaystyle \frac{1}{(2\pi)^{\frac{n}{2}}} \int_{\Qh} \sqrt{\pi}\left(\frac{\mu}{2}\right)^\beta~ t^{\beta}~e^{-\frac{n\sigma^2 t^{2H}}{h^2}}{~}_0\Psi_1
\left[\begin{array}{l|}    \\
	\left(\beta+\frac{1}{2},1\right) 
\end{array} -\frac{\mu^2 t^2d_h(\xi)^2}{4} \right]e^{-i y\cdot \xi}d\xi.\nonumber
%\end{array}
\end{eqnarray}

In particular, in the view of the Mellin inversion formula (\ref{MellinInv}) the identity 
(\ref{compactFH}) becomes then
\begin{eqnarray}
	\label{compactFHMellin}
	%\begin{array}{lll}
	\KH^{\left(\beta\right)}(y,t|\mu,\sigma^2)=\\ \nonumber \\  =\displaystyle \frac{1}{(2\pi)^{\frac{n}{2}}} \int_{\Qh} \left(\frac{1}{2\pi i}\int_{c-i\infty}^{c+i\infty} \mathcal{M}\{f(t)g(t)\}(\omega)~ t^{-\omega}d\omega \right)e^{-i y\cdot \xi}d\xi,\nonumber
	%\end{array}
\end{eqnarray}
with
\begin{eqnarray*}
	f(t):=\sqrt{\pi}\left(\frac{\mu}{2}\right)^\beta t^\beta e^{-\frac{n\sigma^2 t^{2H}}{h^2}} & \& & g(t):={~}_0\Psi_1
	\left[\begin{array}{l|}    \\
		\left(\beta+\frac{1}{2},1\right) 
	\end{array} -\frac{\mu^2 t^2d_h(\xi)^2}{4} \right]
\end{eqnarray*}

In the view of the properties (\ref{MellinP}), (\ref{GammaInt}) and (\ref{WrightFunctionpq}) (see Appendix \ref{FractionalCalcAppendix}) one notice that the functions 
 $f(t)$ and $g(t)$ satisfy the Mellin identities
\begin{eqnarray}
\label{MellinId2k}
\begin{array}{lll}
\mathcal{M}\left\{f(t)\right\}(\omega-s)& =& \displaystyle  \frac{\sqrt{\pi}\left(\frac{\mu}{2}\right)^\beta}{2H}\left(\frac{n\sigma^2}{h^2}\right)^{-\frac{\beta+\omega-s}{2H}}\Gamma\left(\frac{\beta+\omega}{2H}-\frac{s}{2H}\right) \\ \ \\
\mathcal{M}\left\{g(t)\right\}(s)&=&\displaystyle \dfrac{\Gamma\left(\frac{s}{2}\right)}{\Gamma(\beta+\frac{1}{2}-\frac{s}{2})}\left(\frac{\mu^2 t^2d_h(\xi)^2}{4}\right)^{-\frac{s}{2}}.
\end{array}
%\mathcal{M}\left\{t^{\beta-\frac{1}{2}+2k}e^{-\frac{n\sigma^2 t^{2H}}{h^2}}\right\}(s)
\end{eqnarray}

Subsequently, from the Parseval type identity involving the Mellin transform (\ref{MellinParseval}) allows us to represent $\mathcal{M}\{f(t)g(t)\}(\omega)$ as a complex integral over the fundamental strip $\mbox{Re}(s)=c$. In concrete, one has 
\begin{eqnarray*}
\mathcal{M}\{f(t)g(t)\}(\omega)=\\=\frac{1}{2\pi i}\int_{c-i\infty}^{c+i\infty}
\frac{\sqrt{\pi}\left(\frac{\mu}{2}\right)^\beta}{2H}\left(\frac{n\sigma^2}{h^2}\right)^{-\frac{\beta+\omega}{2H}}	\dfrac{\Gamma\left(\frac{s}{2}\right)\Gamma\left(\frac{\beta+\omega}{2H}-\frac{s}{2H}\right)}{\Gamma(\beta+\frac{1}{2}-\frac{s}{2})}\left(\frac{\mu^2 t^2d_h(\xi)^2}{4}\left(\frac{n\sigma^2}{h^2}\right)^{-\frac{1}{H}}\right)^{-\frac{s}{2}}ds.
\end{eqnarray*}

Furthermore, by taking the change of variable $s\rightarrow 2s$ on the above integral, one can recast $\mathcal{M}\{f(t)g(t)\}(\omega)$ as a Wright function of type ${~}_1\Psi_1$. Namely, in the view of Mellin-Barnes representation formula (\ref{WrightFunctionpq}), there holds
\begin{eqnarray}
\label{Mellinfg}
%\begin{array}{lll}
\mathcal{M}\{f(t)g(t)\}(\omega)=\\ =\displaystyle \frac{\sqrt{\pi}\left(\frac{\mu}{2}\right)^\beta}{H}\left(\frac{n\sigma^2}{h^2}\right)^{-\frac{\beta+\omega}{2H}}{~}_1\Psi_1
\left[\begin{array}{l|} \left(\frac{\beta+\omega}{2H},\frac{1}{H}\right)   \\
\left(\beta+\frac{1}{2},1\right) 
\end{array} -\frac{\mu^2 t^2d_h(\xi)^2}{4}\left(\frac{n\sigma^2}{h^2}\right)^{-\frac{1}{H}}\right].\nonumber
%\end{array}
\end{eqnarray}

Thereby, from the previous identity we recognize after a wise change of integration that the function $\KH^{\left(\beta\right)}(y,t|\mu,\sigma^2)$ defined via eq.~(\ref{compactFH}) equals to (\ref{MellinBarnesIntegralH}), concluding in this way the proof of {\bf Theorem \ref{MellinBarnesTheorem}}.
\qed

\section*{Acknowledgement}

The author would like to thank to the anonymous referees for
the careful reading of the paper and for the criticism through the reports. That allowed to improve the quality of the submitted version in a clever style.
\appendix

\section{Fractional Calculus Background}\label{FractionalCalcAppendix}

We aim at presenting in this appendix a systematic account of basic properties and characteristics of generalized Wright functions (also known as Fox-Wright functions (cf.~\cite{MaiP07})) in interplay with the Mellin transform.

\subsection{The Mellin transform}\label{MellinAppendix}

The well-known Mellin transform $\mathcal{M}$ (cf.~\cite{ButJ97}) is defined for a locally integrable function $f$ on $]0,\infty[$ by the integral
\begin{eqnarray}
\label{MellinT}
	\mathcal{M}\{f(t)\}(s)=\int_{0}^{\infty} f(t)t^{s-1}dt, &\mbox{with}& s\in\BC.
\end{eqnarray}

In order to provide the existence of the inverse $\mathcal{M}^{-1}$ of (\ref{MellinT}) through the inversion formula
\begin{eqnarray}
\label{MellinInv}	f(t)=\frac{1}{2\pi i}\int_{c-i\infty}^{c+i\infty}\mathcal{M}\{f(t)\}(s)~t^{-s}~ds, & \mbox{with} & t>0 ~~~\&~~ c=\Re(s)
\end{eqnarray}
in such way that the contour integral is independent of the choice of the parameter $c$, one needs to restrict the domain of analyticity of the complex-valued function $\mathcal{M}\{f(t)\}(s)$ to the fundamental strip $-a<\mbox{Re}(s)<-b$ paralell to the imaginary axis $i\BR$, whereby the parameters $a$ and $b$ are determined through the asymptotic constraint
\begin{eqnarray*}
f(t)=\left\{\begin{array}{lll} 
		O(t^{-a-1}) & \mbox{if} & t\rightarrow 0^+
		\\ \ \\
		O(t^{-b-1}) & \mbox{if} & t\rightarrow \infty
	\end{array}\right..
\end{eqnarray*}  

It is straighforward to see after a wise change of variable on the right hand side of (\ref{MellinT}), we infer that 
\begin{eqnarray*}
	\mathcal{M}\{t^\beta f(t)\}(s)=\mathcal{M}\{f(t)\}\left(s+\beta\right) , & \mbox{for} & \beta \in \BC \\
	\mathcal{M}\{f(t^\gamma)\}(s)=\frac{1}{|\gamma|}(\mathcal{M}f)\left(\frac{s}{\gamma}\right), & \mbox{for} & \gamma\in \BC\setminus \{0\} \\
		\mathcal{M}\{f(\kappa t)\}(s)=\kappa^{-s}(\mathcal{M}f)(s), & \mbox{for} & \kappa>0.
\end{eqnarray*}

With the above sequence of operational identities, neatly amalgamated through the compact formula
\begin{eqnarray}
\label{MellinP}	\mathcal{M}\{t^\beta f(\kappa t^\gamma)\}(s)=\frac{1}{|\gamma|}\kappa^{-\frac{s+\beta}{\gamma}}\mathcal{M}\{f\}\left(\frac{s+\beta}{\gamma}\right)
\end{eqnarray}
carrying the parameters $\beta\in \BC,~\gamma \in \BC\setminus \{0\}$ and $\kappa>0$,
there holds the Mellin convolution theorem 
\begin{eqnarray}
\label{MellinConvolutionM}
\mathcal{M}\{f\star_{\mathcal{M}}g\}(s)=\mathcal{M}\{f\}(s)\mathcal{M}\{g\}(s)
\end{eqnarray}
encoded by the convolution type integral (cf.~\cite[Theorem 3.]{ButJ97})
\begin{eqnarray}
\label{MellinConvolution} (f\star_{\mathcal{M}}g)(t):=\int_{0}^{\infty}f\left(\frac{t}{p}\right)g(p)\frac{dp}{p}.
\end{eqnarray}

We refer to \cite[Section 4.]{ButJ97} for additional properties associated to the Mellin convolution (\ref{MellinConvolution}). In particular, the Parseval type property  
\begin{eqnarray}
\label{MellinParseval} \mathcal{M}\{f(t)g(t)\}(\omega)=\frac{1}{2\pi i}\int_{c-i\infty}^{c+i\infty} \mathcal{M}\{f(t)\}(\omega-s)~ \mathcal{M}\{g(t)\}(s)~ds
\end{eqnarray}
yields straightforwardy from the combination of the set of identities
\begin{eqnarray*}
\mathcal{M}\{f(t)\}(\omega-s)&=&\mathcal{M}\left\{t^{-\omega}f\left(\frac{1}{t}\right)\right\}(s), \\ \mathcal{M}\{f(t)g(t)\}(\omega)&=&\left(t^{-\omega}f\left(\frac{1}{t}\right)\star_{\mathcal{M}}g\right)(1)\end{eqnarray*}
resulting from (\ref{MellinP}) and (\ref{MellinConvolutionM}), respectively, with the set of properties (\ref{MellinConvolution}) and (\ref{MellinInv}).

\subsection{Generalized Wright Functions}\label{GeneralizedWrightSub}

Generalized Wright functions ${~}_p\Psi_q$ are a rich class of analytic functions that include generalized hypergeometric functions ${~}_pF_q$ and stable distributions (cf.~\cite{MaiP07} \& \cite[Chapter 3]{MeerSik11}).
With the aim of amalgamate some the technical work required in subsections \ref{DFPKleinGordonsub} and \ref{GeneralizedWrightSubsection} we will take into account the definition of ${~}_p\Psi_q$ in terms of series expansion
\begin{equation}
\label{WrightSeriespq} {~}_p\Psi_q
\left[\begin{array}{l|} (a_k,\alpha_k)_{1,p}   \\
(b_l,\beta_l)_{1,q} 
\end{array} ~ \lambda \right]=\sum_{m=0}^\infty
\dfrac{\prod_{k=1}^p\Gamma(a_k+\alpha_km)}{\prod_{l=1}^q\Gamma(b_l+\beta_l
	m)}~\dfrac{\lambda^m}{m!},
\end{equation}
where $\lambda\in \BC$, $a_k,b_l\in \BC$ and $\alpha_k,\beta_l\in \BR\setminus\{0\}$ ($k=1,\ldots,p$; $l=1,\ldots,q$).

Here and elsewhere
\begin{eqnarray}
\label{GammaInt}\Gamma(s)=\int_{0}^{\infty} e^{-t}t^{s-1}dt
\end{eqnarray}
stands for the Eulerian representation for the Gamma function.

We note that in particular, that the trigonometric functions may be seen as particular cases of the Mittag-Leffler and Wright functions
\begin{center}
	$
	\displaystyle E_{\rho,\beta}(\lambda)={~}_1\Psi_1
	\left[\begin{array}{l|}  (1,1)  \\
	(\beta,\rho) 
	\end{array}~ \lambda \right]$ resp. $\displaystyle \phi(\rho,\beta;\lambda)={~}_0\Psi_1
	\left[\begin{array}{l|}    \\
	(\beta,\rho) 
	\end{array}~ \lambda \right]$. 
\end{center}

Namely, in view of (\ref{WrightFunctionpq}) and on the Legendre's duplication formula
\begin{eqnarray}
\label{LegendreDuplication}	\Gamma(2s)=\frac{2^{2s-1}}{\sqrt{\pi}}\Gamma(s)\Gamma\left(s+\frac{1}{2}\right)
\end{eqnarray}
one readily has
\begin{eqnarray}
\label{WrighCosine}
%\begin{array}{ccc}
\cos(\lambda)=&{~}_1\Psi_1
\left[\begin{array}{l|}  (1,1)  \\
(1,2) 
\end{array} -\lambda^2 \right]
=&\sqrt{\pi}{~}_0\Psi_1
\left[\begin{array}{l|}    \\
\left(\frac{1}{2},1\right) 
\end{array} -\dfrac{\lambda^2}{4} \right]
%\end{array}
\end{eqnarray}
\begin{eqnarray}
\label{WrighSinc}
%\begin{array}{lll}
\dfrac{\sin(\lambda)}{\lambda}=&{~}_1\Psi_1
\left[\begin{array}{l|}  (1,1)  \\
(2,2) 
\end{array} -\lambda^2 \right]
=&\dfrac{\sqrt{\pi}}{2}{~}_0\Psi_1
\left[\begin{array}{l|}    \\
\left(\frac{3}{2},1\right) 
\end{array} -\dfrac{\lambda^2}{4} \right].
%\end{array}
\end{eqnarray}
showing that $\cos(t)$ and $\frac{\sin(t)}{t}$ are spherical Bessel functions in disguise.

In the paper \cite{KilbasSaigoTrujillo02}, Kilbas et al have checked for $\alpha_k,\beta_l>0$ that ${~}_p\Psi_q$ admits the
the Mellin-Barnes type integral representation
\begin{eqnarray}
\label{WrightFunctionpq} 
\begin{array}{lll}
\displaystyle {~}_p\Psi_q
\left[\begin{array}{l|} (a_k,\alpha_k)_{1,p}  \\
(b_l,\beta_l)_{1,q} 
\end{array} ~ \lambda \right]=\\ \ \\ =\displaystyle \dfrac{1}{2\pi i}
\int_{c-i\infty}^{c+i\infty}
\dfrac{\Gamma(s)\prod_{k=1}^p\Gamma(a_k-\alpha_ks)}{\prod_{l=1}^q\Gamma(b_l-\beta_l
	s)}(-\lambda)^{-s}~ds
\end{array}
\end{eqnarray}
in a way that ${~}_p\Psi_q$ and the inverse of the Mellin transform (see eqs.~(\ref{MellinT}) \& (\ref{MellinInv}) ) are interrelated by the operational formula
\begin{eqnarray*}
	{~}_p\Psi_q
	\left[\begin{array}{l|} (a_k,\alpha_k)_{1,p}   \\
		(b_l,\beta_l)_{1,q} 
	\end{array} ~ \lambda \right]&=&\mathcal{M}^{-1}\left\{\dfrac{\Gamma(s)\prod_{k=1}^p\Gamma(a_k-\alpha_ks)}{\prod_{l=1}^t\Gamma(b_l-\beta_l
		s)}\right\}(-\lambda).
\end{eqnarray*}

This result may be summarized as follows: if intersection between the simple poles $b_l =-m$ ($m\in \BN_0$) of
$\Gamma(s)$ and the simple poles $\frac{a_k+m}{\alpha_k}$
($k=1,\ldots,p;m\in\BN_0$) of $\Gamma(a_k-\alpha_k s)$
($k=1,\ldots,p$) satisfies the condition
$\frac{a_k+m}{\alpha_k}\neq -m$, we have the following characterization:
\begin{enumerate}
	\item In case of $\displaystyle
	\sum_{l=1}^q\beta_l-\sum_{k=1}^p \alpha_k>-1$, the series expansion
	(\ref{WrightSeriespq}) is absolutely convergent for all $\lambda \in
	\BC$.
	\item In case of $\displaystyle
	\sum_{l=1}^q\beta_l-\sum_{k=1}^p \alpha_k=-1$, the series expansion
	(\ref{WrightSeriespq}) is absolutely convergent for all values of
	$|\lambda|<\rho$ and of $|\lambda|=\rho$, $\mbox{Re}(\kappa)>\frac{1}{2}$, with
	\begin{eqnarray*}
		\displaystyle \rho=\displaystyle \dfrac{\Pi_{l=1}^q
			|\beta_l|^{\beta_l}}{\Pi_{k=1}^p |\alpha_k|^{\alpha_k}} & \mbox{and}
		& \kappa=\sum_{l=1}^q b_l-\sum_{k=1}^p a_k+\frac{p-q}{2}.
	\end{eqnarray*}
\end{enumerate}

Other important classes of generalized Wright functions are the modified Bessel functions $$I_{\nu}(u)=\left(\dfrac{u}{2}\right)^\nu{~}_0\Psi_1
\left[\begin{array}{l|}    \\
(\nu+1,1) 
\end{array} ~ \dfrac{u^2}{4} \right]$$ of order $\nu$
and the one-sided L\'evy distribution $L_\nu$ which is represented through the Laplace identity 
 \begin{eqnarray}
\label{LevyDistributions}
\exp(-s^\nu)=\int_{0}^{\infty}e^{-su}L_\nu(u)~du, & 0<\nu<1.
\end{eqnarray}

For the later one we would like to emphasize that $L_\nu$ may be seamlessly described in terms of the Wright functions $\displaystyle \phi(\rho,\beta;\lambda)={~}_0\Psi_1
\left[\begin{array}{l|}    \\
(\beta,\rho) 
\end{array}~ \lambda \right]$ ($-1<\rho<0$) (cf.~\cite{GMainardi98,MaiP07}). 
In concrete, the term-by-term integration of the $k-$terms of $\phi(\rho,\beta;\lambda)$ provided by (\ref{GammaInt}) yields
\begin{eqnarray}
\label{LevyDistributionsWright}
e^{-s^\nu}=\int_{0}^\infty e^{-su} {~}_0\Psi_1
\left[\begin{array}{l|}    \\
(0,-\nu) 
\end{array}~ \frac{1}{u^\nu} \right]\dfrac{du}{u}
\end{eqnarray}
so that (\ref{LevyDistributions}) may be reformulated in terms of the Mellin convolution (\ref{MellinConvolution}). That is, $e^{-s^\nu}=(f\star_{\mathcal{M}}g)(1)$, with
\begin{eqnarray*}
f(t)={~}_0\Psi_1
\left[\begin{array}{l|}    \\
	(0,-\nu) 
\end{array}~ {t^\nu} \right] &\mbox{and} & g(t)=e^{-st}. 
\end{eqnarray*}

Moreover, $L_\nu(u)$ is uniquely determined by 

$$L_\nu(u)=\dfrac{1}{u}{~}_0\Psi_1
\left[\begin{array}{l|}    \\
(0,-\nu) 
\end{array}~ \dfrac{1}{u^\nu} \right].$$

\end{document}